%% file: ms.tex
\documentclass[aps,prb,floatfix,twocolumn,superscriptaddress]{revtex4-2}
\usepackage{amssymb}
\usepackage{amsmath}
\usepackage{graphicx}
\graphicspath{{Figures/}}
\usepackage{float}
\usepackage{braket}
\usepackage{color}
\usepackage{siunitx}
\usepackage[colorlinks=true, linkcolor=blue,  citecolor=blue,
urlcolor=blue]{hyperref}
\usepackage[dvipsnames]{xcolor}

\newcommand{\D}{\mathrm{d}}
\newcommand{\J}{\mathrm{j}}
\newcommand{\LL}{\mathrm{L}}
\newcommand{\RR}{\mathrm{R}}

\begin{document}

\title{{Electrical control of the metal-insulator transition in a one dimensional device}}

\author{J.~Craquelin}
\affiliation{Laboratoire de Physique de l’École normale supérieure, ENS, Université PSL, CNRS, Sorbonne Université, Université Paris Cité, Paris, France.}
\affiliation{Laboratoire de Physique et d'Étude des Mat\'eriaux, ESPCI Paris, PSL University, CNRS, Sorbonne Universit\'e, Paris, France.}

\author{L.~Jarjat}
\affiliation{Laboratoire de Physique de l’École normale supérieure, ENS, Université PSL, CNRS, Sorbonne Université, Université Paris Cité, Paris, France.}
\affiliation{Laboratoire de Physique et d'Étude des Mat\'eriaux, ESPCI Paris, PSL University, CNRS, Sorbonne Universit\'e, Paris, France.}

\author{B.~Hue}
\affiliation{Laboratoire de Physique de l’École normale supérieure, ENS, Université PSL, CNRS, Sorbonne Université, Université Paris Cité, Paris, France.}
\affiliation{Laboratoire de Physique et d'Étude des Mat\'eriaux, ESPCI Paris, PSL University, CNRS, Sorbonne Universit\'e, Paris, France.}
\affiliation{C12 Quantum Electronics, Paris, France.}

\author{A.~Théry}
\affiliation{Laboratoire de Physique de l’École normale supérieure, ENS, Université PSL, CNRS, Sorbonne Université, Université Paris Cité, Paris, France.}
\affiliation{Laboratoire de Physique et d'Étude des Mat\'eriaux, ESPCI Paris, PSL University, CNRS, Sorbonne Universit\'e, Paris, France.}

\author{C.~Fruy}
\affiliation{Laboratoire de Physique de l’École normale supérieure, ENS, Université PSL, CNRS, Sorbonne Université, Université Paris Cité, Paris, France.}
\affiliation{Laboratoire de Physique et d'Étude des Mat\'eriaux, ESPCI Paris, PSL University, CNRS, Sorbonne Universit\'e, Paris, France.}

\author{N.~Struchkov}
\affiliation{C12 Quantum Electronics, Paris, France.}

\author{D.~Stefani}
\affiliation{C12 Quantum Electronics, Paris, France.}

\author{M.M~Desjardins}
\affiliation{C12 Quantum Electronics, Paris, France.}

\author{A.~Cottet}
\affiliation{Laboratoire de Physique de l’École normale supérieure, ENS, Université PSL, CNRS, Sorbonne Université, Université Paris Cité, Paris, France.}
\affiliation{Laboratoire de Physique et d'Étude des Mat\'eriaux, ESPCI Paris, PSL University, CNRS, Sorbonne Universit\'e, Paris, France.}

\author{M.R.~Delbecq}
\thanks{These authors co-supervised this work. e-mail: matthieu.delbecq@phys.ens.fr; takis.kontos@phys.ens.fr}
\affiliation{Laboratoire de Physique de l’École normale supérieure, ENS, Université PSL, CNRS, Sorbonne Université, Université Paris Cité, Paris, France.}
\affiliation{Laboratoire de Physique et d'Étude des Mat\'eriaux, ESPCI Paris, PSL University, CNRS, Sorbonne Universit\'e, Paris, France.}
\affiliation{Institut universitaire de France (IUF), Paris, France}

\author{T.~Kontos}
\thanks{These authors co-supervised this work. e-mail: matthieu.delbecq@phys.ens.fr; takis.kontos@phys.ens.fr}
\affiliation{Laboratoire de Physique de l’École normale supérieure, ENS, Université PSL, CNRS, Sorbonne Université, Université Paris Cité, Paris, France.}
\affiliation{Laboratoire de Physique et d'Étude des Mat\'eriaux, ESPCI Paris, PSL University, CNRS, Sorbonne Universit\'e, Paris, France.}
\affiliation{Institute of Astrophysics, FORTH, GR-71110 Heraklion, Greece}

\begin{abstract}
Controlling the low energy spectrum at the nanoscale with an external physical parameter has become an important resource for quantum devices.
The emergence of an energy gap is one such key feature, linked to the mitigation of decoherence needed for quantum information processing. Indeed, the detrimental effects of high-energy uncontrolled excitations can only be cured at some specific tuning points in general. Achieving an energy gap is a natural way to extend decoherence countermeasures over a finite region of parameter space. This would be particularly useful in view of the recent efforts to build superconducting topological chains in a top-down approach. In this work, we demonstrate a large energy gap by spatially modulating the local potential of a suspended carbon nanotube, exploiting an analogy with condensed matter systems. This gap is homogeneous on the nanotube and tunable by about two orders of magnitude, bringing the electronic system from an insulating state to a near-metallic state at low temperatures.
\end{abstract}

\date{\today}
\maketitle

\section{Introduction}

The metal-insulator transition is an important tracer for condensed matter systems, as it can entail strong correlations or structural changes. It is found in the phase diagram of many exotic materials, such as organic conductors, high-$T_c$ superconductors, and generally complex oxides~\cite{Grandi:20}. For one-dimensional systems or metals with a nested Fermi surface, the Peierls mechanism provides a generic rule for driving a metal into the insulating state. In such a situation, the system minimizes its energy by the emergence of a charge density wave, leading to a spatial modulation of the electronic potential, which opens an energy gap at the Fermi energy.

In mesoscopic systems, it may also be possible to use such a mechanism to engineer a gap to stabilize quantum states. This is particularly relevant for building Kitaev-like chains~\cite{Dvir:23,Haaf:24}. In this context, the top-down approach is expected to simultaneously provide an energy gap and Majorana end states by proper tuning of the local potential. Due to the existence of many sites, the energy gap is expected to be robust against local perturbations, as required for a platform for topologically protected qubits. So far, though, only minimal two-site Kitaev chains have been implemented, leaving open the question of whether a gap can be engineered in mesoscopic devices by spatial control of the local electrical potential. In particular, it is still not yet clear how many sites are needed to enforce a robust energy gap in such chains.

Here, we show that a large and robust energy gap can be induced in a one-dimensional device utilizing a free-standing carbon nanotube over a 15-gate keyboard. The gap is created by spatially modulating the local potential of the multiple gates below the carbon nanotube. The potential is formed by applying alternating voltages, shifted by  $\Delta V_g$ above and below an offset value $\langle V_g \rangle$, from one gate to the next. The gap is tunable over two orders of magnitude, between \SI{200}{\micro\electronvolt} and about \SI{30}{\milli\electronvolt}, as found by transport measurements simply by changing the amplitude of the spatial modulation. In addition to this gap, we also observe conventional charging effects. As such, our system can be viewed as a quantum simulator of a Peierls system with a controlled competition between interactions and incommensurate structural changes.

\section{Results}

\begin{figure}[tph]
{\small \centering\includegraphics[width=0.95\columnwidth,angle=0]{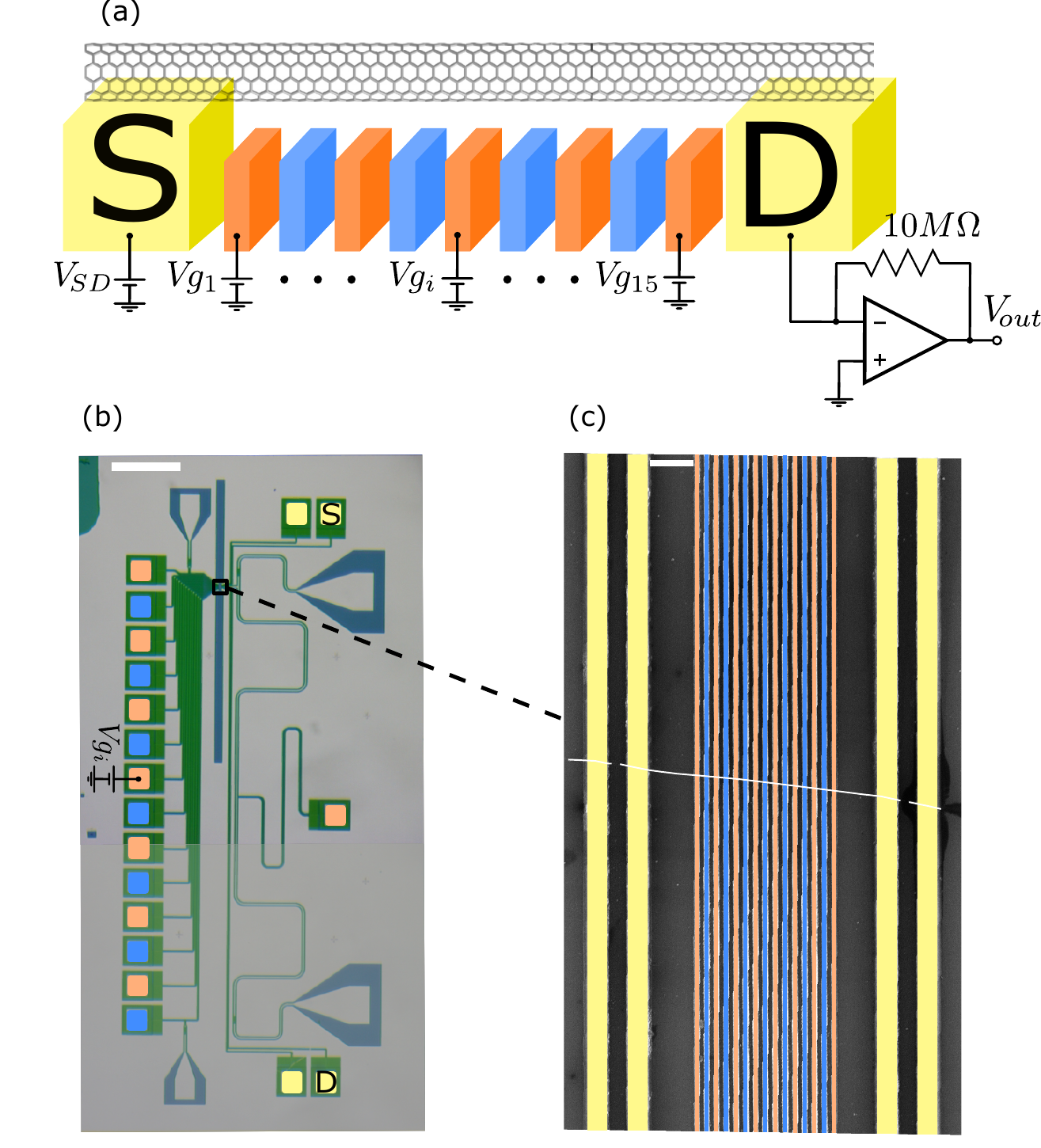}}
\caption{\textbf{Carbon nanotube device}\newline
a. Schematics of the device comprising a carbon nanotube with 15 gates controlled individually. The setup explored here consists of gates with alternating voltage between the "orange" and "blue" gates. The Pd contacts on the carbon nanotube are represented by the yellow source (S) and drain (D) pads. The current through the device is measured with a standard I-V converter amplifier. b. Large scale view of a typical device layout. The device is embedded into a microwave cavity. The bar is \SI{1}{\centi\metre}. c. Close up on a typical device with the 15 (Pd) gates and a suspended carbon nanotube deposited over Pd contacts, in false colors. The bar is \SI{1}{\micro\metre}.}%
\label{fig:device}
\end{figure}

Carbon nanotubes are a priori an interesting platform for the study of the interplay between strong correlations, phonons and local potential perturbations~\cite{Adrian:23,Deshpande:09,Deshpande:08,Ilani:17,Waissman:13,Contamin:23,Desjardins:19}. In particular, interaction induced gaps have been observed in carbon nanotubes in the early works of references~\cite{Deshpande:08,Deshpande:09,Senger:18}. Because interactions are not easily tunable, it has not been possible to extend them in order to demonstrate in-situ, electrically tunable, energy gaps. In addition, many features such as Kondo-like anomalies of unknown origin were observed quite systematically in these measurements. This calls for the use of a different, simpler mechanism to controllably induce an energy gap in one-dimensional systems without relying on electron-electron interactions. This could turn out to be crucial to protect quantum states. More precisely, we focus on a non-interacting mechanism which could ultimately by used to protect quantum states with topological properties, which is a completely different focus from these early studies. Thanks to recent progress in using these 1D conductors in complex circuits, the study of the impact of spatial modulation on their low-energy spectrum is now within reach. Using a nanoscale transfer technique under vacuum developed previously~\cite{Contamin:23}, we present here results on a device that meets the stringent requirements of being suspended over 15 gates, having low contact resistance, and showing weak residual disorder at low temperatures, as suitable for the electrostatic potential to dominate low-energy transport. The corresponding sketch is presented in figure~\ref{fig:device}a. The large-scale view of the entire device is presented in figure~\ref{fig:device}b. Our carbon nanotube device is embedded in a Nb coplanar waveguide cavity which is devised to probe and manipulate ultimately localized topological edge states\cite{Contamin:21}. We focus here on transport measurements to unravel the bulk properties of our device and leave the microwave cavity signals for later studies. A close-up of a typical device is presented in figure~\ref{fig:device}c, in false colors. The \SI{6}{\micro\metre} free-standing part of the nanotube can be seen on the gate electrodes which are located about \SI{150}{\nano\metre} below. Note that for the actual device presented here, the length $L_{NT}$ of the suspended nanotube is about \SI{4}{\micro\metre} due to the slight changes in the layout. The contacts are made of Pd as well as the gate electrodes. The latter choice ensures that we can in-situ check that the nanotube is not collapsing onto the gates (see Methods for details). The current is measured with a standard current-voltage converter, which allows us to extract the differential conductance by means of a numerical derivative. Two important experimental parameters control a priori the transport in our device, besides the individual control of each gate: the offset potential applied to the gates $\langle V_g \rangle$ and the modulation of the potential $\Delta V_g$. The measurements are carried out at $\approx 30 mK$ in a dilution refrigerator.

\begin{figure}[tph]
{\small \centering\includegraphics[width=0.95\columnwidth,angle=0]{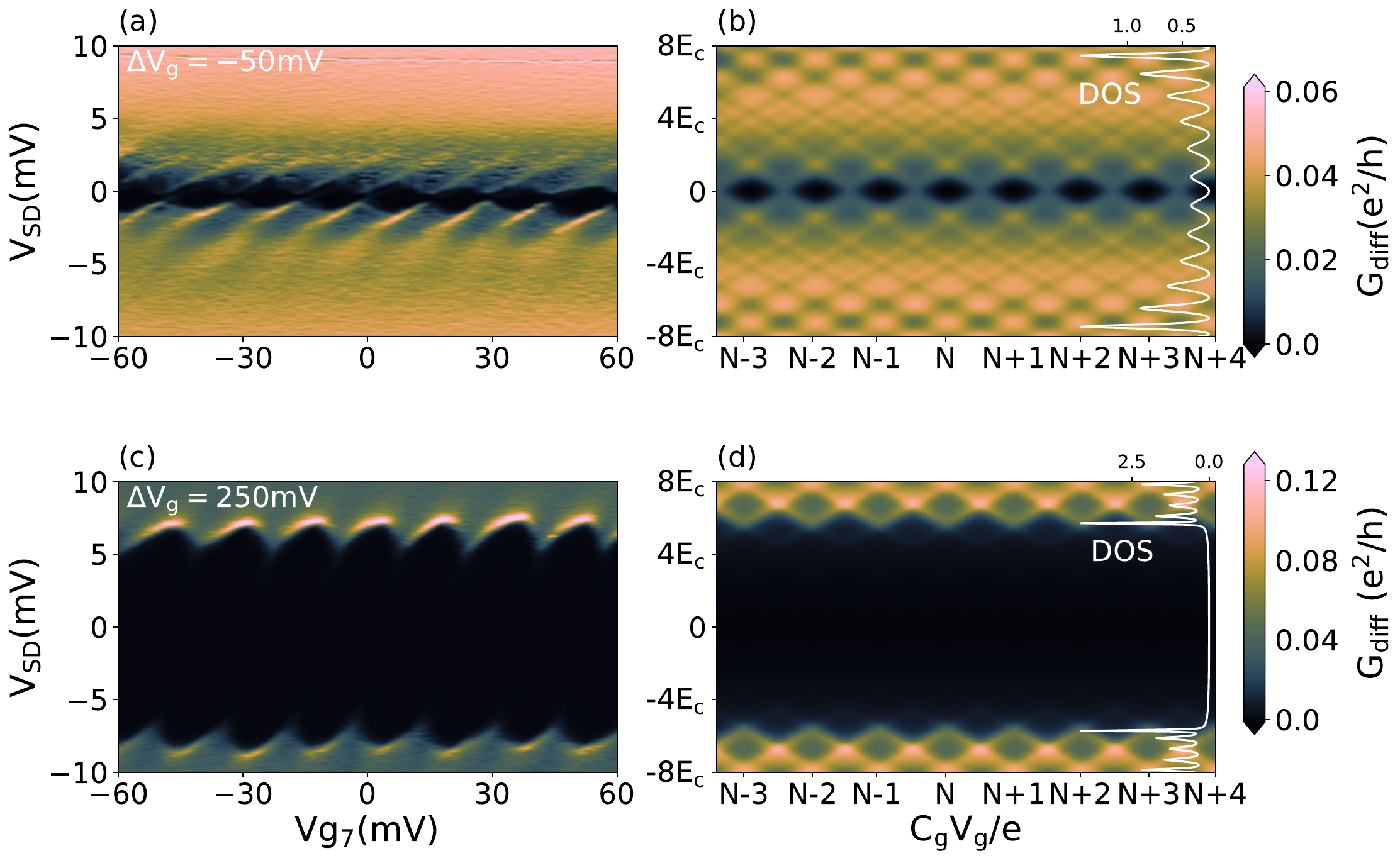}}
\caption{\textbf{Observation of gate induced transport gap}\newline
a. Differential conductance map as a function of $V_{g7}$ for the weak modulation case ($\Delta V_g= \SI{-50}{\milli\volt}$) showing the nearly metallic nanotube case. b. Modeling of the nearly metallic case using the single electron transistor transport theory corresponding to the panel (a.), case of our experiment. c. Differential conductance map as a function of $V_{g7}$ for the strong modulation case ($\Delta V_g= \SI{250}{\milli\volt}$) showing the gapped nanotube case. d. Modeling of the gapped case using the single electron transistor transport theory corresponding to the panel (c.), case of our experiment. All the experimental conductance maps were realized at $\braket{V_g} = \SI{0.2}{\volt}$. The vertical cuts for the theoretical density of states are shown in white solid lines in panels b and d.}%
\label{fig:gap}
\end{figure}

The transport properties of our system are presented in figure~\ref{fig:gap}a for a small modulation amplitude $\Delta V_g=\SI{-50}{\milli\volt}$ and for an offset gate value of $\braket{V_g} = \SI{0.2}{\volt}$. Here, a single gate ($V_{g7}$) is swept while the others are kept constant in order to ensure no major change in the disorder configuration while keeping a overall gating effect. One observes the conventional Coulomb blockade pattern in color-scale plots. Coulomb diamonds allow us to extract a charging energy $E_c \approx \SI{1.25}{\milli\electronvolt}$. The pattern, with blockaded regions below the diamonds and featureless regions above the diamonds, is similar to that of a single-electron transistor (SET). This is fully consistent with the expected level spacing $h v_F/2L_{NT} \lesssim \SI{0.45}{\milli\electronvolt} $, where $v_F \approx 8.10^{5} \SI{}{\metre\per\second\tothe{1}}$ is the Fermi velocity of metallic single-wall carbon nanotubes. The situation changes strikingly when we raise $\Delta V_g$ to \SI{250}{\milli\volt}. Instead of Coulomb diamonds, we observe a large stripe of very weak conductance, signaling an energy gap $E_g$. There are residual modulations of the unblocking threshold that are reminiscent of the charging effects previously observed. From the conductance map presented in figure~\ref{fig:gap}c, we can estimate $E_g \approx \SI{5}{\milli\electronvolt}$. The fact that this gap is a single particle property and not an interaction-driven suppression of transport can be examined through a theoretical modeling of the observed conductance maps. Figures 2b and 2d present such a modeling for the two experimental situations of Figures 2a and 2c. The model used to produce these maps is simply that of a SET (see the Methods section and the Supplementary for details). In conventional SETs, the density of state of the blockaded island is assumed to be constant~\cite{Korotkov:94}. However, it is possible to include a textured density of states $\mathcal N(\epsilon)$, including a gap $E_g$. The generic case of the ``semiconductor model'' implies a Dirac spectrum with $\mathcal N (\epsilon)=|\epsilon|/\sqrt{\epsilon^2-E_{g}^2}$ but one can also include the theoretically expected density of states using tight binding method and Green's function techniques. In that case, the input parameter is the imaginary part of the total retarded Green's function of the chain (see Supplementary): $\mathcal N_{chain} (\epsilon)=-\frac{1}{\pi} \Im m[\mathcal G^R(\epsilon)]$. Although the simple ``semiconductor model'' density of states can account for our experimental data, we have used here $\mathcal N_{chain} (\epsilon)$ for the density of states input in figure~\ref{fig:gap}b and 2d, anticipating on the discussion of the actual physical mechanism at work here. Similarly to a quantum dot contacted with superconducting electrodes~\cite{Bruhat:16}, the effect of the gap is to shift the Coulomb diamonds to higher bias, seemingly truncating the Coulomb blockade pattern at low bias. This shows that the increase of the modulation amplitude has the effect of driving the system from a metallic state to essentially an insulating state with a sizable gap.

The possibility to induce an energy gap by a spatial modulation of the electronic potential in a one-dimensional system has regained interest recently in the context of topological systems~\cite{Dalibard:18}. In particular, a chain with a suitable modulation pattern can be used to engineer a Su--Schrieffer--Heeger (SSH) chain in the non-superconducting case~\cite{Dalibard:18}, or, when combined with proximity-induced superconductivity in an appropriate configuration, a Kitaev chain~\cite{Dvir:23,Haaf:24}. The system we have in mind is the SSH chain which can host exotic Jackiw-Rebbi edge states~\cite{Gangadharaiah:12} of fractional charge. Such a topological setup has also been shown to be the basis for non-abelian braiding~\cite{Klinovaja:13}, alike their superconducting counterpart potentially hosting Majorana modes. The topological protection of such a quantum computing scheme would arise from the possibility to control efficiently a large bulk gap. Obtaining a large and electrically controllable bulk gap is the focus of the present study. While we implement the simplest gate modulation setting here which is not exactly the SSH regime, our circuit is designed along the general efforts to control and isolate topological excitations. In addition, since it is embedded in a microwave cavity, it is clearly designed in the perspective to detect and braid topological edge excitations in an SSH-like setup~\cite{Contamin:21}.

The robustness of the energy gap is linked to the periodic setting of the spatial modulation, like in a conventional lattice. In more complex condensed matter systems, a gap can arise thanks to the Peierls mechanism both in the case of commensurate or incommensurate modulations with the underlying lattice. The latter case requires special features of the Fermi surface such as nesting, leading to the emergence of a charge density wave, which in turn induces an insulating state. The insulating state can arise from other mechanisms than the charge density wave, e.g. electronic correlations~\cite{Grandi:20}. In addition, the charge density wave state may not be stable enough to prevail. In our case, we induce a synthetic charge density wave owing to our local gate electrodes. Obtaining a truly one-dimensional system is in addition very difficult in devices in practice, particularly in a top-down approach, due to the fabrication-induced disorder. Disorder is here the main competing mechanism to prevent the controlled opening of an energy gap from electrical spatial modulations of the electron potential. It is thanks to the use of nanoscale assembled single wall carbon nanotubes that we can investigate such a controllable setup experimentally here.

\begin{figure*}[tph]
\centering
\includegraphics[width=0.95\linewidth,angle=0]{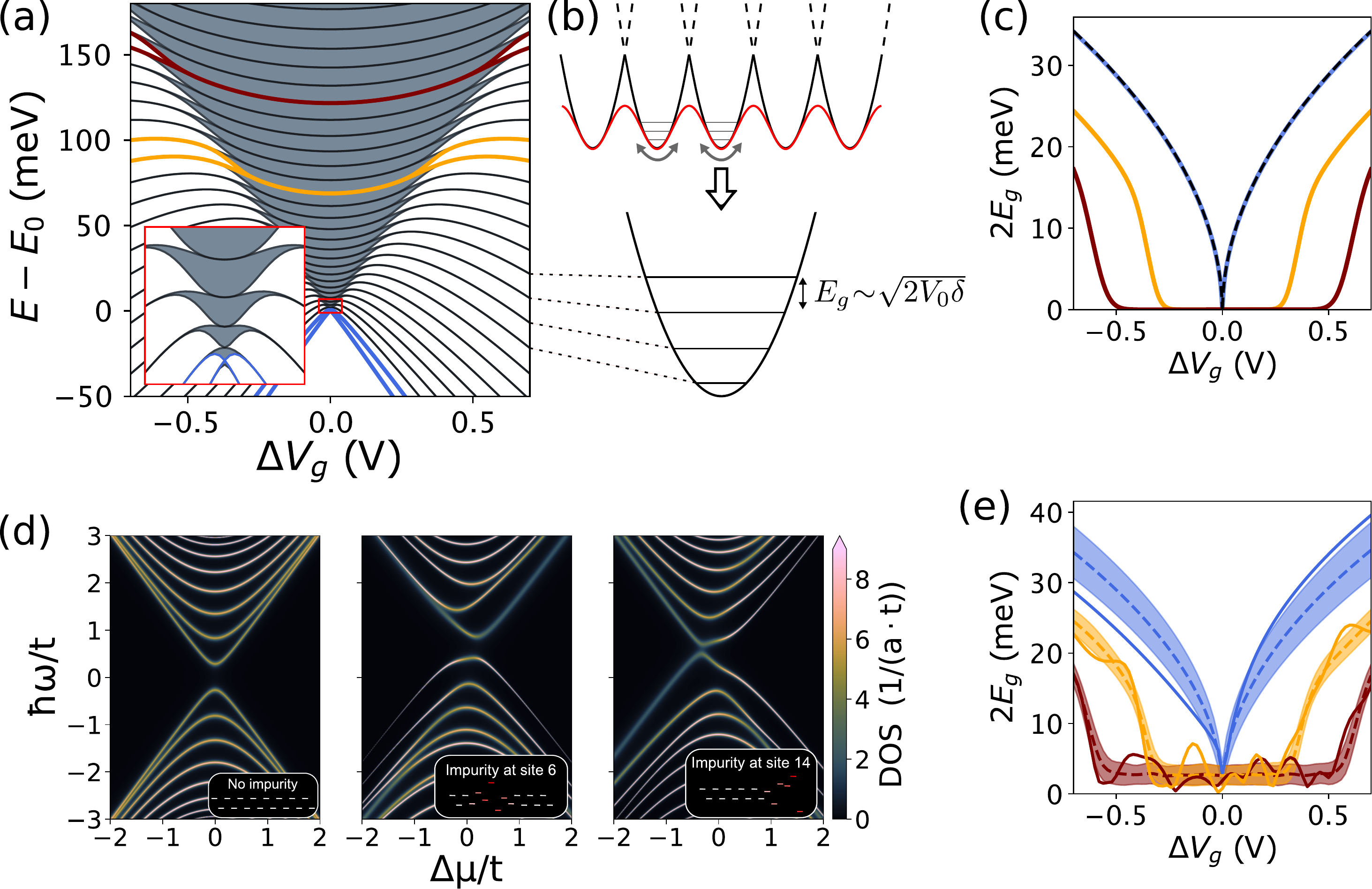}
\caption{\textbf{Metal-Insulator transition physics of the device.}
(a) Numerically calculated energy spectrum obtained by solving the Schrödinger equation with an infinite cosine potential of amplitude $\Delta V_g$, assuming a parabolic dispersion relation with intrinsic semiconductor band gap $E_0$. The inset provides a zoomed-in view of the band bottom. The grey shaded areas are the region of continuum of states.
(b) Illustration of the cosine modulation potential (red curve), showing the quadratic approximation valid near each potential minimum. In the strong modulation limit, the bands become increasingly narrow and converge toward the discrete energy levels of a harmonic oscillator.
(c) Energy gaps between successive bands as a function of modulation amplitude $\Delta V_g$, shown for various chemical potentials measured from the band bottom. The colors match the corresponding bands depicted in panel (a). The black dashed line correspond to the asymptotic Mathieu solution in the large modulation limit.
(d) Calculated density of states (DOS), given in units of $1/(a t)$, where $a$ is the lattice spacing and $t$ is the hopping parameter, for a chain consisting of $N = 16$ sites with alternating chemical potentials $\Delta \mu$ (see supplementary material). The left panel illustrates the ideal case without disorder. The middle and right panels show the impact of impurities introduced at sites $i=6$ and $i=14$, respectively, modifying the local chemical potential at these sites and their neighbors. The chemical potential shift experienced at each other site $j$ decays exponentially as $t\,e^{-(i-j)^2/\sigma^2}$, with $\sigma=3$.
(e) Same calculation as panel (a), now incorporating electrostatic disorder (details provided in supplementary material). Dashed lines represent average results over multiple disorder realizations, shaded regions correspond to one standard deviation around these averages, and solid lines depict specific individual disorder realizations.}
\label{fig:cartoon}
\end{figure*}

The theoretical problem of generating exactly a gap in a one dimensional system with a cosine potential was considered as early as 1952 by Slater~\cite{Slater:52}, as a complementary study of Kronig and Penney. The interest of the cosine potential is also that the problem in 1D can be solved exactly through a mapping to the Mathieu functions. The spectrum directly stems from the Mathieu characteristic function~\cite{Slater:52,Cottet:02,Wilkinson:18} and is shown in figure~\ref{fig:cartoon}a. As originally found by Slater, bands are formed as soon as there is a finite modulation amplitude $\Delta V_g$. Interestingly, for very large amplitude, the system consists of localized harmonic oscillators which display a harmonic spectrum, as sketched in figure~\ref{fig:cartoon}b which shows that the quadratic bottom of the potential in each modulation becomes dominant. This analogy with harmonic spectra enables one to understand the dependence of the gap as a function of the oscillation amplitude in particular. This equation is now also heavily used in the Josephson junction literature~\cite{Cottet:02,Wilkinson:18}. For the first band, one can obtain explicit analytical formulae as found in the literature~\cite{Wilkinson:18}. Anticipating on the discussion on our device, we consider here realistic values for the gate modulations amplitude $|\Delta V_g|$. For the first gap, in the small modulation amplitude, one expects a gap $E_g \approx 2 |\Delta V_g|$ whereas in the large modulation amplitude $E_g \approx 2\sqrt{2\delta |\Delta V_g|}$ with $\delta=\hbar^2 \pi^2/2m_{\rm eff} \lambda^2$, $m_{\rm eff}$ being the effective mass of the electrons and $\lambda$ being the modulation wavelength. The large $\Delta V_g$ limit has a gap with a $\sqrt{|\Delta V_g|}$ dependence, as a direct consequence of the harmonic spectrum. The expected dependence of the gap delimiting the different bands is shown in figure~\ref{fig:cartoon}c for 3 particular gaps illustrating the mechanisms, corresponding to different chemical potentials away from the bottom of the band.  Interestingly, while the square root dependence can be identified for $|\Delta V_g|>0.5V$, we find that there can be a threshold for the onset of the gap depending on the band considered. This is going to be important for the discussion on our data.

Before analyzing the data, it is important to have an idea of the inevitable effect of disorder, even if it is weak. We can tackle numerically this by two means: first, we can add a random potential to the Mathieu equation and solve the corresponding modified Mathieu equation as done in the supplementary ; second, we can also express the chain problem using a tight binding hamiltonian and calculate the density of state from the imaginary part of the retarded Green's function. The latter is shown in figure~\ref{fig:cartoon}d for no disorder, disorder in the middle and disorder on the side. In this case, the finite size of the system implies that the gap does not fully close for $\Delta V_g=0$ because it is limited by the level spacing of the full size chain. In addition, owing to the single fermionic level of each chain site, the gap opening has a different functional dependence as a function of $\Delta \mu/t \sim \Delta V_g$. The result of disorder in this case is twofold: first the gap dependence becomes asymmetric both horizontally and vertically; second, it generates 'bumps' in the spectrum. This becomes even clearer if one considers how the bands of figure~\ref{fig:cartoon}a change with disorder (see Supplementary). The main effect of disorder is to induce random fluctuations in the dependence of the gap as a function of  $\Delta V_g$. This yields fluctuations in the gap curve as shown in figure~\ref{fig:cartoon}e, as a consequence of randomness.  Interestingly, this is reminiscent of the quantum Hall effect where the Landau levels are broadened by disorder although they are primarily harmonic oscillators.

We now compare the predictions of the Mathieu spectrum with our experimental findings. By solving numerically the Laplace equation of the potential generated by the gate setting (see Supplementary for more details), we can show that although the electrodes only take two distinct values $\braket{V_g}\pm \Delta V_g$, it can be considered a good approximation, mainly because the nanotube is at a distance comparable to the pitch of the gate 'keyboard' (see Supplementary for details) from the gate plane. Leaving the full ``Poisson-Schrödinger'' problem for later studies, we now focus on the spectrum of electrons in such a simplified setup.
Taking an effective mass of $m_{\rm eff} \approx 1\times10^{-32} \SI{}{\kilo\gram}$ (i.e., a semiconducting gap of \SI{40}{\milli\electronvolt} with $v_F = 8\times 10^{5}\SI{}{\metre\per\second\tothe{1}}$) and $\lambda\approx \SI{400}{\nano\metre}$, we have $\delta\approx \SI{100}{\micro\electronvolt}$ (see the Methods section for details). From the Coulomb blockade maps, one can extract an average lever arm of $\bar{\alpha} = 0.12$. With these values, our system is in the large modulation limit, and for $\Delta V_g = \SI{250}{\milli\volt}$ (as it is the case in figure~\ref{fig:gap}b), it gives a gap of $E_g \approx \SI{6}{\milli\electronvolt}$ close to $E_g \approx \SI{5.5}{\milli\electronvolt}$ found in the experiment, even for the first gap function corresponding to blue dashed line in figure~\ref{fig:cartoon}c. We refine the comparison with the Mathieu equation prediction in the next section.

\section{Discussion}

\begin{figure*}[tph]
\centering
\includegraphics[width=0.95\linewidth,angle=0]{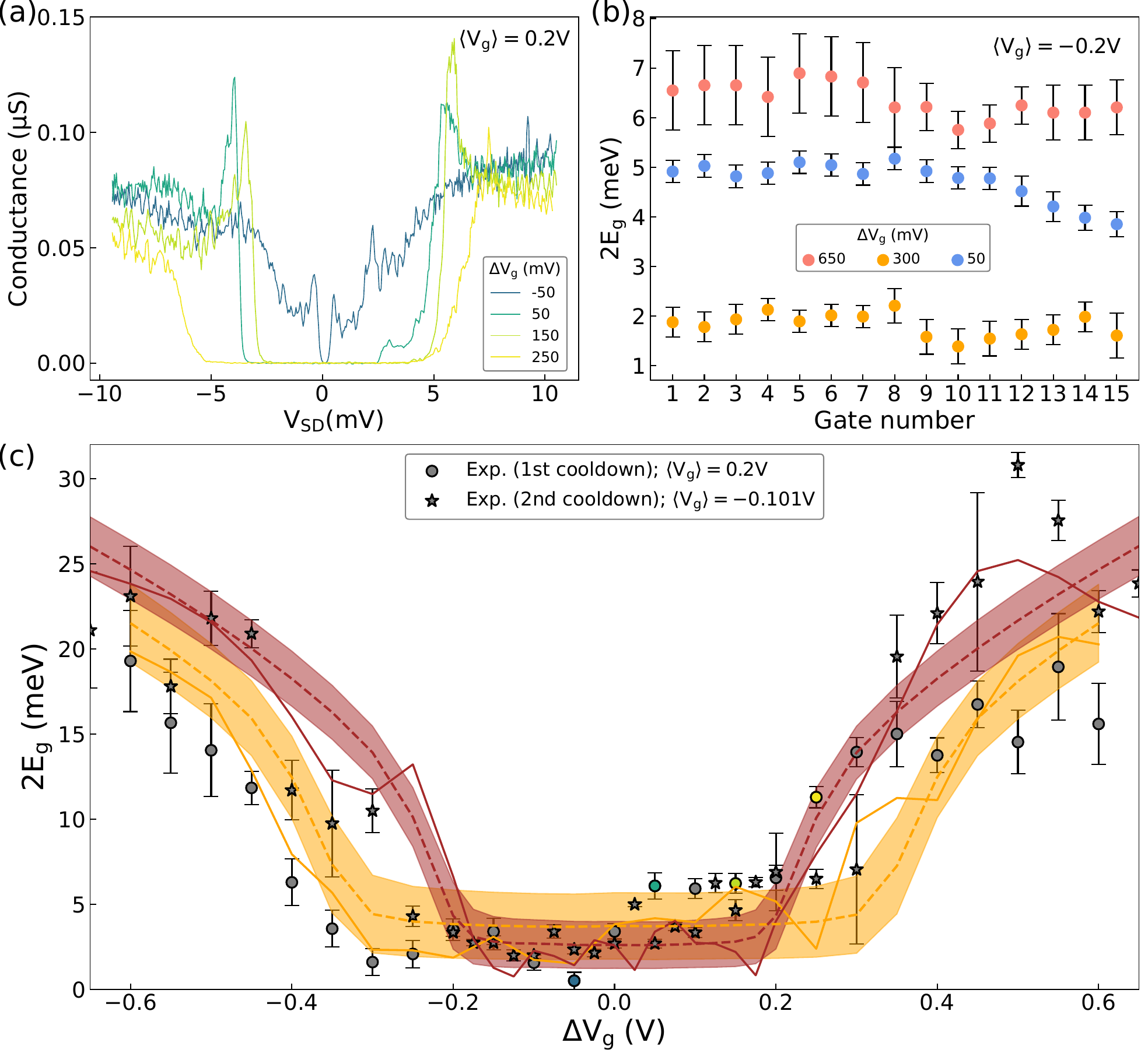}
\caption{\textbf{Electrical control of the gap}\newline
a. Differential conductance as a function the bias voltage $V_{SD}$ highlighting the electrical control of the energy gap using the spatial modulation, for different values of $\Delta V_g $. In these linear plots, one should bear in mind that there is also a contribution from the charging energy, hence the asymmetry with respect to zero bias in general. b. Dependance of the extracted energy gap as a function of the gate number for 3 different gap values. c. Dependence of the energy gap (measured as described in figure~\ref{fig:gap}) on the gate spatial modulation. Circle and star symbols represent two distinct sets of measurements performed on the same sample during separate cooldown cycles. The offset gate voltage, $\langle V_g \rangle$, was adjusted to maintain a similar chemical potential relative to the semiconducting band gap. Colored markers indicate linear cuts shown in panel (a). The error bars correspond to the systematic error linked to the determination of the bias threshold for the conductance onset. Corresponding numerical simulations are shown in orange and brown; dashed lines represent the average values obtained from multiple disorder realizations, shaded areas indicate one standard deviation from these averages, and solid lines correspond to specific disorder realizations that qualitatively match the experimental data.}%
\label{fig:tuning}
\end{figure*}

We can further test the above picture by systematically studying how the energy gap measured depends on $\Delta V_g$. First, it is instructive to show individual conductance curves in order to examine the "hardness" of the obtained gap, which plays an important role for quantum devices. As shown in figure~\ref{fig:tuning}a, the conductance ratio between low and high biases is larger than two orders of magnitude, which places our system in the "hard gap" regime. Note here that the conductance curves are asymmetric with respect to positive/negative bias $V_{SD}$, because of charging effects.
Variations in the amplitude of the gap can be tested locally for different gate electrodes. In a situation where the electrodes are in a $\braket{V_g} \pm\Delta V_g$ configuration, one can sweep each electrode individually around its original value to probe how the gap can be potentially affected by each gate individually. If the system is homogeneous, we expect only a weak change. If the gate is close to a disorder site, we expect a sizable response of the nanotube. We present such a study in figure~\ref{fig:tuning}b (see also Supplementary Figure 3). One observes that the gap is homogeneous up to $15 \%$ along the nanotube, which means that for a given $\Delta V_g$, the gap extracted for each gate has on average a $15\%$ difference from the mean value of the gap at that $\Delta V_g$. More data on the gap spatial homogeneity is shown in the Supplementary figure 3. To compensate these small fluctuations, we average the gap over 2 gates and we turn to its dependence on $\Delta V_g$. The corresponding plot is presented in figure~\ref{fig:tuning}c. This panel contains measurements from two different cooldowns. At large negative $\Delta V_g$, the gap is large, of about \SI{30}{\milli\electronvolt}. Then, the gap reaches a plateau for $|\Delta V_g|\lesssim $\SI{0.3}{\volt} for the first cool down and \SI{0.2}{\volt} for the second. In a first cool down, a minimum of about \SI{200}{\micro\electronvolt} around $\Delta V_g = \SI{-50}{\milli\volt}$ is reached. For positive $\Delta V_g$, the gap grows again, in a rather symmetric fashion up to about \SI{30}{\milli\electronvolt} again.
The gap function versus $\Delta V_g$ from the Mathieu equation, taking into account of a constant gate lever arm of $0.25$ is overlaid  on our experimental data for the two different cooldowns. The dashed lines are the predictions for the disorder-averaged gap between bands 6 and 7 (red) and 8 and 9 respectively (yellow). The continuous lines are the predictions for a particular realization of the disorder potential. The shaded area is the 1 sigma region away from the mean. We find a good agreement between theory and experiment for the two cooldowns. The stripes around the theoretical curves correspond to the expected spread arising from disorder. During the first cool-down, we also could test the gap opening and closing for other offset gate voltage. The corresponding measurements are shown in the supplementary. One can check that they are fully consistent with figure~\ref{fig:tuning}c, even though the onset of the increase of the gap can happen for a lower $|\Delta V_g|$. This calls for the study of the electrical of the gap for different$\braket{V_g}$.

\begin{figure*}[tph]
{\small \centering\includegraphics[width=0.95\linewidth,angle=0]{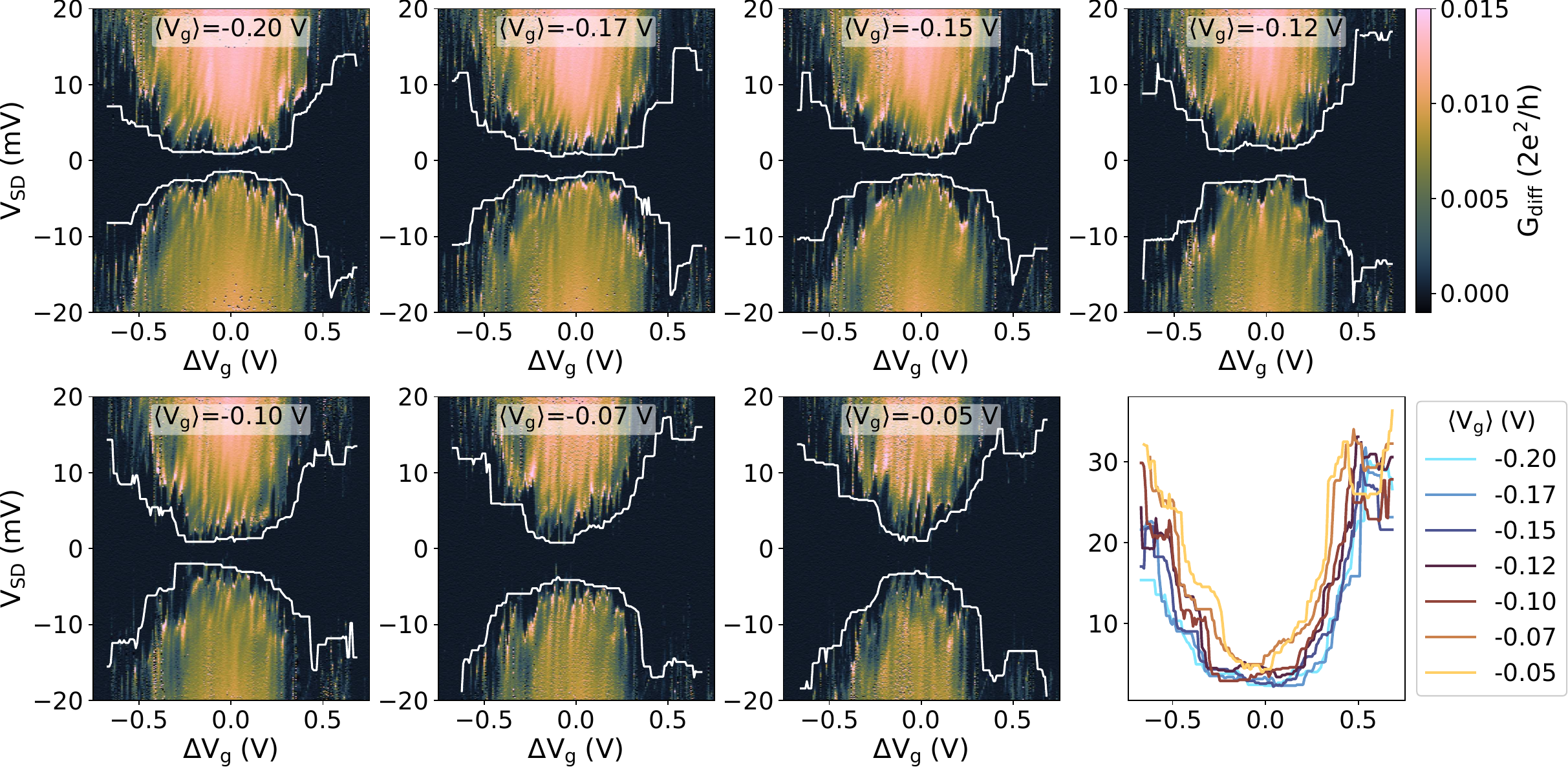}}
\caption{\textbf{Opening and closing the gap for different chemical potentials}\newline
Evolution of the depressed conductance region as a function of gate modulation amplitude $\Delta V_G$ for different offset gate voltages $\braket{V_g}$ (indicated at the top of each panel) done in the second cooldown. In each plot, the white line marks the conductance threshold that defines the depressed region. The bottom-right panel summarizes the amplitude of the resulting conductance gap in bias voltage as a function of $\braket{V_g}$.}%
\label{fig:robsutness}
\end{figure*}

The effect of different $\braket{V_g}$ is shown in Figure~\ref{fig:robsutness} which corresponds to data from the second cooldown. Here, instead of determining the gap from local Coulomb diamond pattern like we have done so far, we present directly the differential conductance data in colorscale plots as a function of $V_{SD}$ and $\Delta V_g$ for different $\braket{V_g}$ ranging from \SI{-0.20}{\volt} to \SI{-0.05}{\volt}. The dark region delimited by the white lines allows us to read off the gap edge (at this scale the Coulomb diamonds are too small to be visible but they are always present). We observe a roughly symmetric gap opening and closing with a flat part as in figure~\ref{fig:tuning}c, consistently for all the values of offset gate voltage. Interestingly, this flat part extends less for $\braket{V_g}$ closer to the bottom of the band as summarized in the lower right panel. This is fully consistent with the Mathieu spectrum shown in figure~\ref{fig:cartoon} and was already appearing in figure~\ref{fig:tuning}c. Indeed, as one moves away from the bottom of the band, one opens a gap between higher index bands, which have a gap onset for higher $|\Delta V_g|$.

How many gates are needed for a robust gap to develop~? Intuitively, the larger the number of gates, the more robust the gap should be. However, the harmonic picture which should prevail for large $|\Delta V_g|$ shows that one could also simply increase the depth of the potential. This situation is however equivalent to a single trap which is not the setup we have in mind especially in connection to topologically protected quantum states. In order to address the question of the number of gates needed in our setup, we construct site by site the chain as shown in Figure~\ref{fig:bottomup}. In the main text, we start by the middle and in the supplementary, we start by the side of the chain. The corresponding potential landscape is shown above each panel. In the upper left panel, we show the effect of a single central gate. In this case, we see Coulomb diamonds which do not change in size for negative $\Delta V_g$ and are larger around $\Delta V_g\approx$ \SI{0.50}{\volt}. We see no clear gap opening. Such a measurement, as well as the similar one when a single side gate is used instead of the middle gate is interesting to rule out potential effects from spurious uncoupled quantum dots. When single gates are actuated, we observe qualitatively different conductance features, which shows that the number of gates matters and that we do not reveal with our measurements the existence of uncontrolled dots in our setup or a property of the nanotube itself such as its intrinsic gap which is far away from our offset gate setting (see supplementary). In the lower right panel, we show the effect of the full gate keyboard. Up to 5 gates in the center of the chain, we see low conductance regions but no clear gap opening and closing. A clear gap opening appears however for 7 gates modulated and above. The shape of the gap versus $\Delta V_g$ remains essentially unchanged up to the weak disorder induced fluctuations (see supplementary for details).

\begin{figure*}[tph]
{\small \centering\includegraphics[width=0.95\linewidth,angle=0]{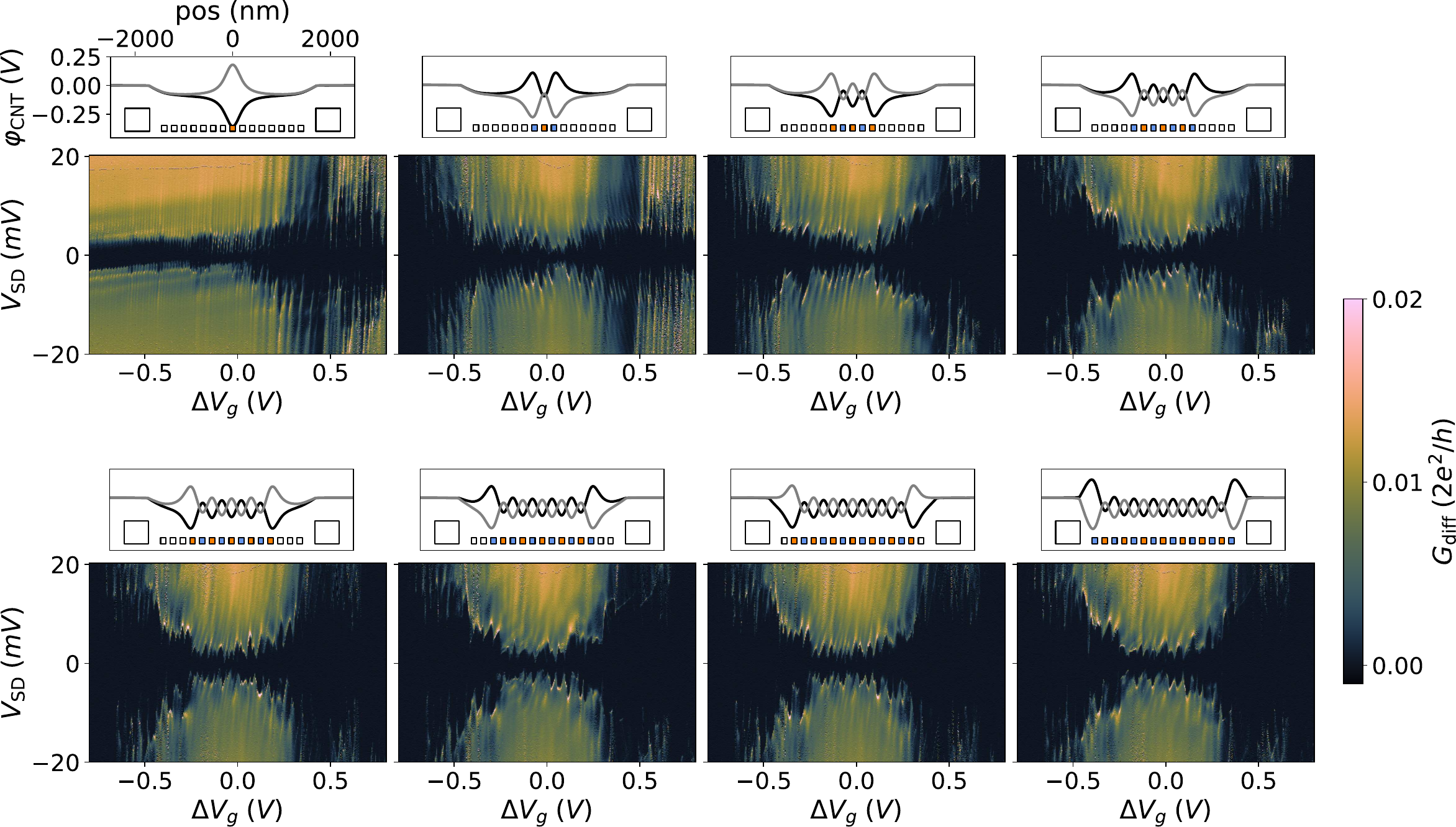}}
\caption{\textbf{Site by site construction of the gap}\newline
Construction of the chain exhibiting the gap site by site at a value of $\braket{V_g}=\SI{-0.101}{\volt}$ (second cooldown), split in eight panels. The upper left panel corresponds to a single site whose gate voltage is being modulated. The lower right corresponds to the full chain studied in the main text. Above each conductance plot, a schematic illustrates the corresponding modulated gates, in blue and orange (alternating gates), while white squares indicate non-modulated gates. Large squares at the edges represent source and drain contacts. Grey and black curves show the calculated electrostatic potential profile along the CNT for the two extreme modulation amplitudes.}%
\label{fig:bottomup}
\end{figure*}

In conclusion, we have presented a one-dimensional device, based on a carbon nanotube, with an engineered energy gap stemming from spatial modulations of the electronic potential. These modulations, which can be tuned simply by local gate voltages, directly control the amplitude of the energy gap. Our method can be generalized to any low-dimensional system and may be relevant to implement one-dimensional chains with exotic end states~\cite{Gangadharaiah:12}. Finally, since our system is naturally coupled to phonons or photons, one of the potential applications of our findings is the quantum simulation of fermion-boson problems and, in particular, the Peierls instability~\cite{Adrian:23} or the braiding of non-abelian excitations using cavity photons \cite{Contamin:21}.

\section{Methods}

\subsection{Device fabrication}

The circuit is made up of the following elements: 15 DC gates (12 DC only, 2 RF + DC and 1 cavity gate), 1 microwave CPW cavity with its central conductor capacitively coupled to the nanotube, 2 electrodes (source and drain), 2 electrodes used to cut the nanotube and finally 2 trenches to enable the CNT deposition.
The DC lines, RF lines, ground plane and cavity are etched by reactive ion etching (RIE) after a full chip metalization of $90 \mathrm{nm}$ of niobium. The cavity is a centimetric CPW resonator $L = \lambda/2$. The impedance of all the DC and RF lines is $50 \Omega$ to avoid reflections with external RF and DC sources.
The 15 DC gates are \SI{100}{\nano\metre} wide and spaced by \SI{100}{\nano\metre} from each other and are made of a stack of \SI{30}{\nano\metre} Ti / \SI{25}{\nano\metre} Pd. The electrodes are made of a bilayer of \SI{200}{\nano\metre} Ti / \SI{25}{\nano\metre} Pd. The gate-CNT distance of \SI{170}{\nano\metre} is chosen so that the measurements are not sensitive to the spatial displacement of the CNT~\cite{Waissman:13}. Finally, since the tube extends over a long distance ($\sim \SI{4}{\micro\metre}$), the electrodes must be high enough so that the tube does not collapse onto the gates.
The trenches are $\sim \SI{15}{\micro\metre}$ deep and are etched by RIE while the rest of the chip is protected by a thick $\sim \SI{3}{\micro\metre}$ resist.

CNTs are grown on silicon cantilevers by $\mathrm{CH_4}$ chemical vapor deposition. The nanotubes are imaged using a scanning electron microscope.
Once the circuit is fabricated, a nanotube is deposited on top of the contact electrodes under vacuum through a nanoscale transfer method. This process allows us to select a clean CNT that never undergoes any clean-room process.

A CNT is brought close to the chip surface while a \SI{2}{\volt} DC bias is applied between the contact electrodes under a high $10^{-7} \SI{}{\milli\bar}$ vacuum. When the nanotube reaches the surface of the electrodes, a current is measured, allowing us to detect it electrically. Before removing the comb, the tube is cut on both sides by applying a high current between the lead and the cut electrode (typically $5-15\SI{}{\micro\ampere}$). This last step also acts as an annealing procedure. Finally, one can measure the source-drain resistance that typically reaches $ \gtrsim \SI{500}{\kilo\ohm}$.\newline

\subsection{Modeling of transport using the SET theory}

We consider a model analogous to the single-electron transistor (SET)~\cite{Korotkov:94}. The nanotube is considered an island with a charging energy $E_c$ with a density of states $\rho_{CNT}(E)$. The assumption of continuum density of state is justified by the length of the nanotube leading to a small level spacing as well as the broadening of the energy levels. When $\Delta V_g$ varies, the nanotube goes from metallic to insulator, resulting in a gap $E_g$ opening in its density of states. We take a generic gapped density of states:
\begin{equation}\label{eq:dos_CNT}
    \rho_{CNT}(E) = \Re e \left(\frac{E + i\eta}{\sqrt{(E+i\eta)^2 - E_{g}^2}}\right)
\end{equation}
with $E$ the energy, $\eta \rightarrow 0^+$ being a small inelastic phenomenological broadening.

We consider a series of tunneling events in which electrons can tunnel between the leads and the island. These processes are treated perturbatively and coherent effects are neglected, which leads to a semi-classical sequential regime. In such a case, one can consider the tunneling rates $\Gamma_\mathrm{L/R}^+(N)$ associated with an electron going from a lead to the island and $\Gamma_\mathrm{L/R}^-(N)$ when an electron goes from the island to a lead. These rates depend on the total number $\mathrm{N}$ of excess electrons on the island.

The current flowing through the device can be expressed as :

\begin{widetext}
\begin{equation}
    I = -e\sum_{N=0}^{\infty}\left[\Gamma_\LL^+(N) - \Gamma_\LL^-(N) \right]P_N = -e\sum_{N=0}^{\infty}\left[\Gamma_\RR^-(N) - \Gamma_\RR^+(N) \right]P_N
\end{equation}
\end{widetext}
where $P_N$ is the stationary probability of having $N$ electrons on the carbon nanotube.

The stationary probability distribution satisfies, in addition to the conservation of probabilities:
\begin{equation}
    P_N \left[ \Gamma_\LL^+(N) + \Gamma_\RR^+(N)  \right] = P_{N+1} \left[\Gamma_\LL^-(N+1) + \Gamma_\RR^-(N+1) \right]
\end{equation}

The derivation of the tunneling rates $\Gamma_\mathrm{L/R}^\pm(N)$ is done in detail in the Supplementary.
\begin{widetext}
\begin{equation}
    \Gamma_{\LL/\RR}^{\pm}(N) = 2 \pi \gamma_L \int \D E \rho_{CNT}(E)
     \left(1-n_f(E, T)\right) \times n_f\left(\pm E_c\{1+2N- 2\frac{C_gV_g}{e} + \frac2e(V_{\LL/\RR} - V_{\RR/\LL})C_{\RR/\LL}\} \pm E, T\right)
\end{equation}
\end{widetext}
with $n_f(E,T)$ being the Fermi-Dirac distribution at temperature $T$,
The conductance maps of figure~\ref{fig:gap}c and 2d have been produced with the above expressions, using the following parameters.

\begin{widetext}
\begin{center}
\begin{tabular}{ c | c | c | c | c | c | c | c }

$E_c (meV)$ & $C_L (aF)$ & $C_R(aF)$ & $C_g(aF)$ & $k_BT(\mu eV)$ & $\gamma_L(\mu eV)$ & $\gamma_R(\mu eV)$ & $\eta(\mu eV)$   \\

\hline

 1.25 & 2.1 & 2.1 & 2.1 & 167 & 2 & 1 & 20 \\

\end{tabular}
\end{center}
\end{widetext}

where $eV_L(meV)$, $eV_R(meV)$ and $eV_g(meV)$ belong to $[-\SI{10}{\milli\electronvolt}, \SI{10}{\milli\electronvolt}]$,$[-\SI{10}{\milli\electronvolt}, \SI{10}{\milli\electronvolt}]$ and $[\SI{1.25}{\milli\electronvolt}, \SI{25}{\milli\electronvolt}]$ respectively. The thermal broadening is assumed to be large here in order to account for electron heating which is inevitable at very large bias. This does not affect the conductance owing to the large gaps observed.\newline

{\small {\noindent\textbf{Data availability } The data that support the findings of this work are available from the corresponding authors upon request.}}

{\small {\noindent\textbf{Code availability } The codes that support the findings of this work are available from the
corresponding authors upon request.}}

{\small 
}

\medskip
{\noindent{\small \textbf{Acknowledgements. } We thank L. De Medici and D. Roditchev for fruitful discussions. Fundings: This work was supported by the French National Research Agency (ANR) MITIQ (T.K.), the ANR JCJC STOIC (ANR-22-CE30-0009) (M.R.D.), the BPI project QUARBONE (T.K.), by the ANR through the France 2030 programme through the PEPR MIRACLEQ (ANR-23-PETQ-0003) (M.R.D.), the QRADES Quantera project (T.K. and A.C.) and the DarkQuantum ERC project (T.K. and A.C.).}}
\newline\newline
{\small {\noindent\textbf{Authors contributions }
MRD and TK designed and supervised the experiment. JC fabricated the device and performed the measurements. JC did the modelling with inputs from AC, MRD and TK. LJ, BH, AT and CF contributed to the data taking and the data acquisition setup. NS, DS and MMD provided the carbon nanotube material and contributed to the nano-assembly setup. TK wrote the manuscript with inputs from all the authors. MRD and TK co-supervised this work and contributed equally as co-last authors; either name may be listed last when citing this contribution.
\newline\newline
{\small {\noindent\textbf{Competing interests. } Authors affiliated with C12 Quantum Electronics have financial interest in the company. T.K. and M.R.D. declare equity interest in C12 Quantum Electronics. JC, LJ, BH, AT, CF and AC declare no competing interests.}}

\clearpage
\onecolumngrid
\begin{center}
\textbf{\Large Supplementary materials for ``Electrical control of the metal-insulator transition in a one dimensional device''}
\end{center}

\setcounter{equation}{0}
\setcounter{section}{0}
\setcounter{figure}{0}
\setcounter{table}{0}
\setcounter{page}{1}

\renewcommand{\theequation}{S\arabic{equation}}
\renewcommand{\thesection}{S\arabic{section}}
\renewcommand{\bibnumfmt}[1]{[S#1]}
\renewcommand{\citenumfont}[1]{S#1}

\renewcommand{\figurename}{Supplementary Figure}
\renewcommand{\tablename}{Supplementary Table}

\input{supplementary}

\makeatother

\end{document}

%% file: supplementary.tex
\section{Tight binding modelling of the chain: density of states}
In the main text, we argue that when we apply a voltage modulation $\Delta V_g$, it opens a gap in the metallic band of the CNT. To capture this, we can consider a discrete chain of N dots where each site has a staggered chemical potential $\mu_i = \braket{\mu} \pm \Delta \mu$ and is coupled to its neighbors with tunneling energy $t_i$. The Hamiltonian of such a chain reads: 

\begin{equation}
    \hat{H}_{chain} = \sum_i \mu_i c^{\dagger}_ic_i + \sum_i t_ic^{\dagger}_{i+1}c_i + h.c
\end{equation}

For a chain with periodic boundary conditions (or infinite chain), this solves exactly and the eigen energies are given by: 

\begin{equation}\label{eq:tight_binding_infinite}
    E_n(k) = \pm \sqrt{\Delta\mu^2 + \left(2t \cos{k_n/2}\right)^2}
\end{equation}

With $k_n = \frac{2n \pi}{N} \in [-\pi, \pi ]$. The chain is metallic for $\Delta \mu = 0$ and has a gap $\propto \lvert\Delta\mu\rvert$ for $\lvert\Delta\mu\rvert \ne 0$. These properties remain valid when the system is coupled to fermionic reservoirs. However, their manifestation must be analyzed using the Green's function formalism which properly accounts for the open-system nature and non-equilibrium effects. Charging effects can also be straightforwardly included in the constant interaction model scheme in S1 since it conserves the total number of charges.

The hamiltonian of the reservoir (source and drain electrodes) is described by a quasi continuum of modes.

\begin{equation}
    \hat{H}_{bath} = \sum_{k, O \in \{S, D\}} \varepsilon_{O, k} c^{\dagger}_{O, k} c_{O, k} \; \; \; \; \mathrm{and } \; \; \; \; \\
    \hat{H}_{int} = \sum_{k}t_S c^{\dagger}_1 c_{S, k} + t_D c^{\dagger}_N c_{D, k} + h.c
\end{equation}

The standard one time retarded Green's function is given by:

\begin{equation}
    \mathcal{G}^{R}_{B, A} = -i \theta(t) \braket{\{B(t), A(0)\}}
\end{equation}

Here, $\theta$ denotes the Heaviside step function and $\{ \}$ is the anti-commutation operator acting on A and B. The Green's function $\mathcal{G}^R(t)$ acting on all pairs of operators $c_i, c_j, c^{\dagger}_i, c^{\dagger}_j$ can be represented by a (2N, 2N) matrix which is built with the following blocks: 

\begin{equation}
    \mathcal{G}^{R}_{i, j} = 
    \begin{pmatrix}
        \mathcal{G}^{R}_{c_i, c^{\dagger}_j} & \mathcal{G}^{R}_{c_i, c_j} \\
        \mathcal{G}^{R}_{c^{\dagger}_i, c^{\dagger}_j} & \mathcal{G}^{R}_{c^{\dagger}_i, c_j}
    \end{pmatrix}
\end{equation}

In the frequency domain, it is given by:

\begin{equation}
    \mathcal{G}^{R}(\omega) = 
    \begin{pmatrix}
        \hbar\omega - \mu_1 + i \Gamma_S/2  & -t_1^{*} & 0 & 0 & \dots \\
    
        -t1 & \hbar\omega - \mu_2 &  -t^{*}_2 & 0 & \dots\\
        0 & -t_2 &  \hbar\omega - \mu_3 & -t^{*}_3 & \ddots \\
        \ddots & \ddots & \ddots & \ddots & \ddots 
    \end{pmatrix}^{-1}
\end{equation}

Where $\Gamma_{S/D} = 2 \pi \lvert t_{S/D} \rvert^2 \rho_{S/D}(\omega)$ is the rate at which the chain can exchange electrons with the bath. The density of states can be computed site by site:
\begin{equation}
    \mathcal N_{chain} (\epsilon) = - \frac{1}{\pi} \mathcal{I}m\operatorname{Tr}\left[\mathcal{G}^R(\epsilon) \right] 
\end{equation}

It is important to note that, even in the continuous limit $N \rightarrow{\infty}$ this model cannot be mapped directly onto the Mathieu problem in which the potential is cosinusoidal. In fact, the staggered potential leads to a band gap that can not be quantitatively compared with the one extracted from the Mathieu equation. However, this model also provides a convenient framework to introduce disorder and to investigate its impact, particularity on the stability of the gap.

Interestingly, we see that the density of states calculated here displays a minimum gap linked to the level spacing of the full chain. This is consistent with the minimum gap of about \SI{200}{\micro\electronvolt} observed in the first cooldown. Of course, this is a lower bound which is in general increased by the disorder.

\section{Transport calculation details}\label{section:transport_calc_details}
\subsection{Description of the model}
We consider a model analogous to the single-electron transistor (SET). We rederive here the standard SET theory \cite{korotkov_intrinsic_1994} in the regime where, instead of a metallic island, the metallic dot (the nanotube in our case) acts as an element with a tunable density of state. The nanotube is coupled to a left and right electrodes through tunnel junctions with capacitance and resistance $C_{\mathrm{L/R}}$ and $R_{\mathrm{L/R}}$, as shown in supplementary Figure~\ref{fig:schematic_theory}.
If the electronic temperature is low enough $T\ll e^2/C_j$ while the tunnel resistances are larger than the quantum unit $R_\mathrm{j} \gg R_\mathrm{Q} = \hbar/2e^2$, the current through the SET can be controlled by varying the gate voltage $V_g$ and thus the number of electrons on the island.

\begin{figure}[h]
    \centering
    \includegraphics[width=0.65\textwidth]{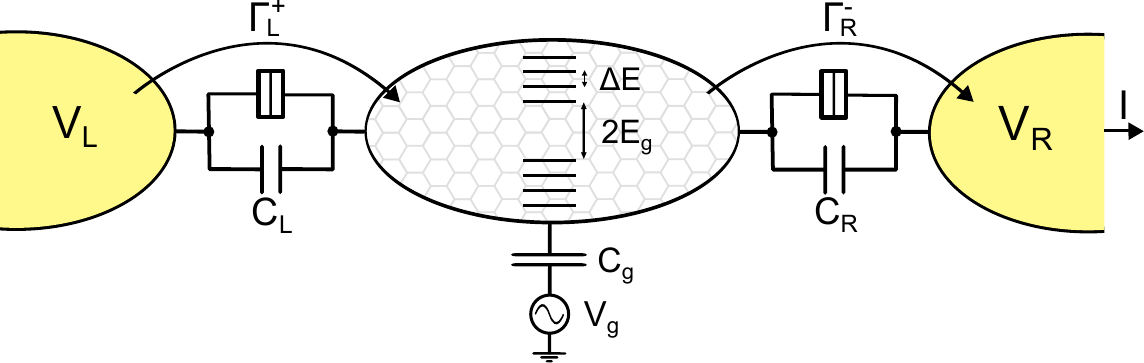}
    \caption{\textbf{Schematics of the modeled device.} 
    A voltage is applied between the two electrodes ($V_L$ and $V_R$). The two tunnel junctions rates are $\Gamma_\mathrm{L/R}^{+/-}(N)$. In the middle, a carbon nanotube is gated in such a way that its density of states can be controlled continuously from metallic to insulator associated to a gap $E_g$. The energy levels spacing is labeled $\Delta E$.}
    \label{fig:schematic_theory}
\end{figure}

\subsection{Sequential tunneling regime}
We define the tunneling rates $\Gamma_\mathrm{L/R}^+(N)$ associated with an electron going from a lead to the island and $\Gamma_\mathrm{L/R}^-(N)$ when an electron goes from the island to a lead. These rates depend on the total number $\mathrm{N}$ of excess electrons on the island. \cite{bruus_many-body_2004}

In such a framework, the electron dynamics is described by a master equation :
\begin{equation}
    \frac{\D P_N(t)}{\D t} = \sum_{\J = \LL/\RR} \left\{ \Gamma_\J^+(N-1)P_{N-1}(t) + \Gamma_\J^-(N+1)P_{N+1}(t) - \left[\Gamma_\J^+(N) + \Gamma_\J^-(N)\right]P_N(t) \right\}
\end{equation}

where $P_N(t)$ is the probability of having N electrons on the island.

The stationary probability distribution satisfies :
\begin{equation}
    P_N \left[ \Gamma_\LL^+(N) + \Gamma_\RR^+(N)  \right] = P_{N+1} \left[\Gamma_\LL^-(N+1) + \Gamma_\RR^-(N+1) \right]
\end{equation}

And the conservation of charges :

\begin{equation}
    \sum_{N = 0}^{\infty}P_N = 1
\end{equation}

In addition, in this stationary regime, the current that goes from left to center is equal to the current that goes from center to right.
\begin{equation}
    I_{\mathrm{st}} = \langle I_\LL(t) \rangle_{\mathrm{st}} = \langle I_\RR(t) \rangle_{\mathrm{st}}
\end{equation}
Which leads to
\begin{equation}
    I_{\mathrm{st}} = -e\sum_{N=0}^{\infty}\left[\Gamma_\LL^+(N) - \Gamma_\LL^-(N) \right]P_N = -e\sum_{N=0}^{\infty}\left[\Gamma_\RR^-(N) - \Gamma_\RR^+(N) \right]P_N
\end{equation}

\subsection{Computation of the tunnel rates}
To compute the tunnel rates, we will consider the following perturbative tunneling Hamiltonians :
\begin{equation}
    H_{T, \LL} = \sum_{\nu_{\LL}, \nu_D} t_{\LL,\nu_{\LL}, \nu_D}c_{\nu_\LL}^\dagger c_{\nu_D}  + h.c
\end{equation}

\begin{equation}
    H_{T, \RR} = \sum_{\nu_{\RR}, \nu_D} t_{\RR,\nu_{\RR}, \nu_D}c_{\nu_\RR}^\dagger c_{\nu_D}  + h.c
\end{equation}

where $H_{T, \LL}$ (resp. $H_{T, \RR}$) is the tunneling Hamiltonian from the left (resp. right) lead to the dot. $t_{\LL,\nu_{\LL}, \nu_D}$ (resp. $t_{\RR,\nu_{\RR}, \nu_D}$) is the hopping term associated with an electron going from the dot (state $\nu_D$) to the lead (state $\nu_{\LL/\RR}$). These states are associated with the fermionic operators $c_{\nu_{\RR / \LL}}^\dagger $ and $c_{\nu_{\RR / \LL}}$.
The transition rate $\Gamma_\LL^+(N)$ from a state $\ket{i_N}$ to a state $\ket{f_{N+1}}$ where 1 electron goes from the left lead to the dot is given by the Fermi's Golden Rule:

\begin{equation}\label{FGR_Gamma_L+}
    \Gamma_\LL^+(N) = 2\pi\sum_{i_N, f_{N+1}} \lvert\bra{f_{N+1}}H_{T, \LL}\ket{i_N}\rvert^2 W_{i_N}\\
    \times \delta(E_{f_{N+1}} - E_{i_{N}})
\end{equation}
With $\ket{f_{N+1}} = c_{\nu_D}^\dagger c_{\nu_\LL} \ket{i_N}$.
$W_{i_N}$ is a thermal distribution function.
When the dot goes from a states with $N$ electrons to a state with $N+1$ electrons (or the other way around), the energy variation of the circuit includes three different contributions : the variation of kinetic energy of the electrons $\Delta E_{kin}^{\pm}(N)$, the variation of electrostatic energy $\Delta E_{el}^{\pm}(N)$ and the work $W_{\LL/\RR}^{\pm}$ associated to the lead losing an electron.
To evaluate the electrostatic part, one needs to consider the so called constant interaction model:
\begin{equation}
    H_{CI} = E_c\left( \sum_{\nu_D} n_{\nu_D}\right)^2
\end{equation}
associated with energy
\begin{equation}
    E_{el}^+(N) = E_c N^2
\end{equation}

\begin{equation}
    \Delta E_{el}(N)^{\pm} = E_c(1\pm2N)
\end{equation}

The variation of kinetic energy is given by the energy gained by the electron from leaving an orbital $\nu_\LL$ to go to an orbital $\nu_D$:
\begin{equation}
    \Delta E_{kin}^{\pm}(N) =\pm (\varepsilon_{\nu_D} - \varepsilon_{\nu_{\LL/\RR}})
\end{equation}

And the work of the lead \cite{korotkov_intrinsic_1994} :
\begin{equation}
    W_{\LL / \RR}^{+/-} = \pm e\frac{C_gV_g}{C_\Sigma} \pm e(V_{\RR / \LL} - V_{\LL / \RR})\frac{C_{\RR/\LL}}{C_\Sigma}
\end{equation}
This allows to compute to total energy variation of the system when an electron tunnels in or out of the dot:
\begin{equation}
    \Delta E_{tun}^{\pm}(N) = \Delta E_{el}(N)^{\pm} + \Delta E_{kin}^{\pm}(N) - W_{\LL / \RR}^{\pm}
\end{equation}
In the case of an electron going from left to the dot, equation (\ref{FGR_Gamma_L+}) becomes :

\begin{equation} \label{eq:FGR2}
     \Gamma_\LL^+(N) = 2\pi\sum_{\nu_D, \nu_\LL}\sum_{i_N} \lvert t_{\LL,\nu_{\LL}, \nu_D} \rvert^2
     \lvert\bra{i_{N}}c_{\nu_\LL}^\dagger c_{\nu_D}c_{\nu_D}^\dagger c_{\nu_\LL}\ket{i_N}\rvert^2 W_{i_N}
     \delta\left(\Delta E_{el}(N)^{+} + \Delta E_{kin}^{+}(N) - W_{\LL}^{+} \right)
\end{equation}

Considering that the lead and the dot are independent, we can write:
\begin{equation}
    \sum_{i_N} \lvert \bra{i_{N}}c_{\nu_\LL}^\dagger c_{\nu_D}c_{\nu_D}^\dagger c_{\nu_\LL}\ket{i_N}\rvert^2 W_{i_N} =
    \left( \sum_{i_{N,D}} \lvert \bra{i_{N, D}} c_{\nu_D}c_{\nu_D}^\dagger \ket{i_{N,D}}\rvert^2 W_{i_{N,D}} \right)
    \left( \sum_{i_{L}}\lvert  \bra{i_{\LL}} c_{\nu_\LL}c_{\nu_\LL}^\dagger \ket{i_\LL}\rvert^2 W_{i_\LL} \right)
\end{equation}

$ \bra{i_{N, D}} c_{\nu_D}c_{\nu_D}^\dagger \ket{i_{N,D}} $ is either equal to 0 or 1, as it gives the occupation of the orbital $\nu_D$ in the initial state $\ket{i_{N,D}}$. In consequence, these sums can be replaced by the Fermi-Dirac distribution functions: $\left(1-n_f(\varepsilon_{\nu_D}, T)\right)n_f(\varepsilon_{\nu_\LL}, T)$

Equation (\ref{eq:FGR2}) becomes:

\begin{equation}
\begin{split}
    \Gamma_\LL^+(N) = 2 \pi \int \int \D\varepsilon_D \D\varepsilon_\LL \rho_{CNT}(\varepsilon_D) \rho_\LL (\varepsilon_\LL)
    \lvert t_{\LL,\nu_{\LL}, \nu_D} \rvert^2 \left(1-n_f(\varepsilon_{\nu_D}, T)\right)n_f(\varepsilon_{\nu_\LL}, T)\\
    \times \delta\left(\Delta E_{el}(N)^{+} + \Delta E_{kin}^{+}(N) - W_{\LL}^{+} \right)
    \end{split}
\end{equation}

We can consider that around the fermi energy, the density of states of the lead $\rho_\LL$ and $t_{\LL,\nu_{\LL}, \nu_D}$ are constant whereas the density of states for the nanotube has a gap $E_g$:

\begin{equation}\label{eq:dos_CNT}
    \rho_{CNT}(E) = \Re\left(\frac{E + i\eta}{\sqrt{(E+i\eta)^2 - E_g^2}}\right)
\end{equation}

with $E$ the energy, $\eta \rightarrow 0^+$ being a small inelastic phenomenological broadening. This is a simplified density of states which can be used. In the Figure~2 of main text, we use rather the full density of states calculated in the previous section.

and finally:

\begin{equation}
    \Gamma_\LL^+(N) = 2 \pi \gamma_L \int \D\varepsilon_D  \rho_{CNT}(\varepsilon_D)
     \left(1-n_f(\varepsilon_{D}, T)\right)n_f(E_c\{1+2N- 2\frac{C_gV_g}{e} + \frac2e(V_\LL - V_\RR)C_\RR\} + \varepsilon_{D}, T)
\end{equation}

For a metallic dot, this formula can be further simplified integrated (\cite{bruus_many-body_2004}, \cite{korotkov_intrinsic_1994}) and becomes the well-known formula from which the Coulomb blockade peaks arise.

\begin{equation}
    \Gamma_{\LL/\RR}^{+/-}(N) =  f(\Delta E_{tun, \LL/\RR}^{+/-}) \quad  \quad \mathrm{and} \quad \quad f(E) = \frac{E}{e^{E/k_bT}-1}
\end{equation}

\newpage

\section{Additional data}

\subsection{Transport map as a function of $\braket{V_g}$}

We show here a transport map from cooldown~2 of the conductance map $G_{diff}$ as a function of $\braket{V_g}$ and $V_{SD}$ at $\Delta V_g=\SI{0}{\volt}$. We see Coulomb diamonds that do not close at low bias and with a larger separation as $\braket{V_g}$ is increased towards the semi-conducting gap. We interpret this conductance gap as due to electrostatic disorder which induces a transport gap even at zero modulation, as detailed in~\ref{ssec:numerical_mathieu}.
\begin{figure}[h]
    \centering
    \includegraphics[width=0.9\textwidth]{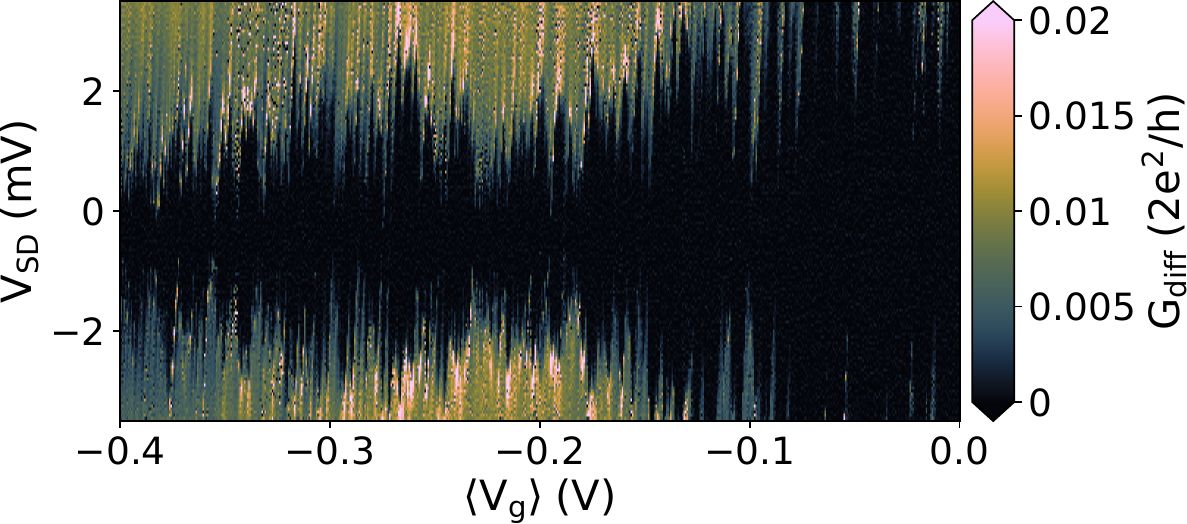}
    \caption{\textbf{Large offset gate voltage scan. }
    Conductance map $G_{diff}$ as a function of $\braket{V_g}$ and $V_{SD}$ at $\Delta V_g=\SI{0}{\volt}$ for cooldown~2.}
    \label{fig:Vsd_vs_Vgavg}
\end{figure}

\subsection{Spatial dependence of the gap}
\begin{figure}[h]
    \centering
    \includegraphics[width=0.5\textwidth]{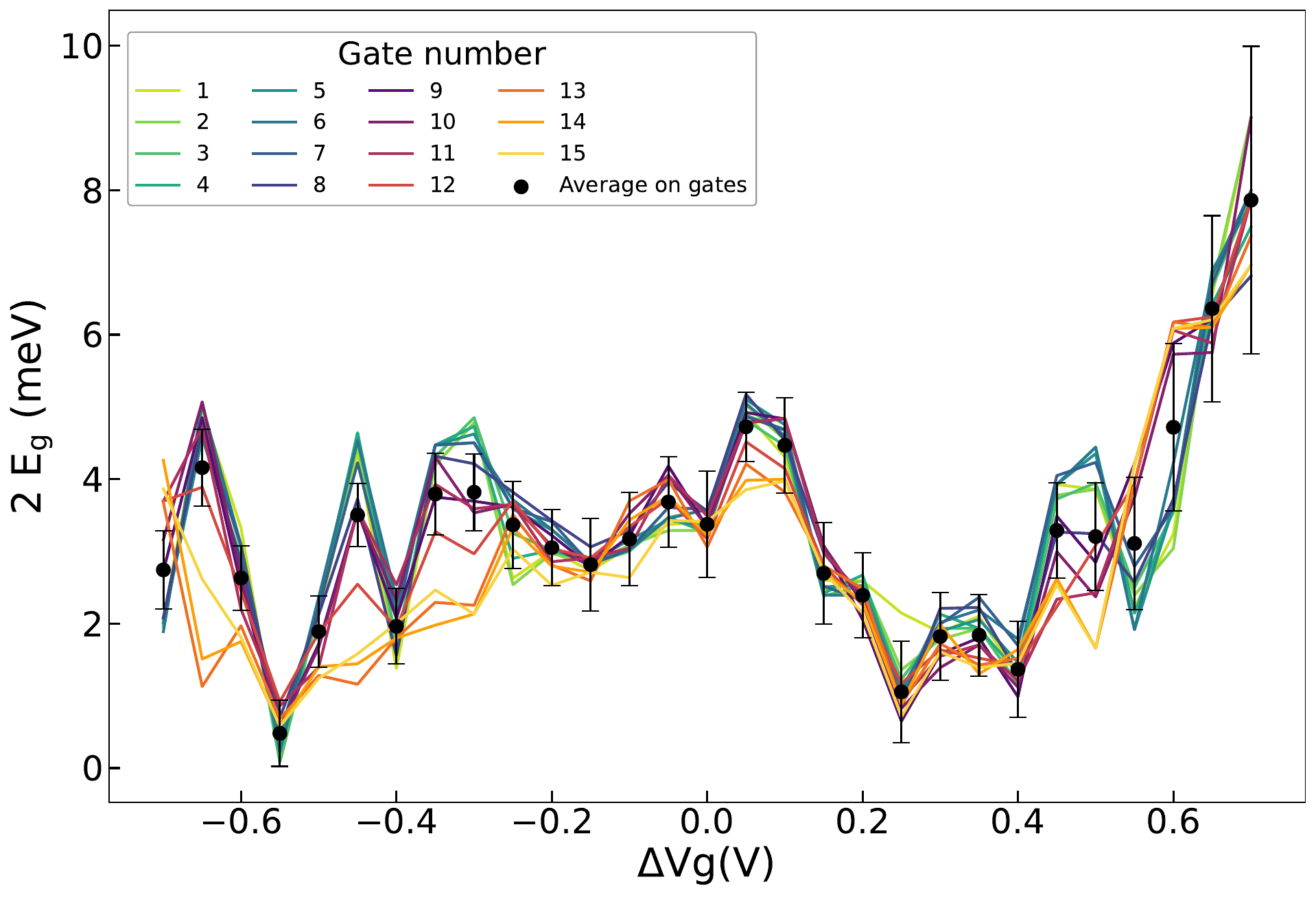}
    \caption{\textbf{Additional data for local gate response.} 
    Gap in mV versus gate modulation $\Delta V_g$ in V, taken at $\braket{V_g} = \SI{-0.2}{\volt}$. Each data point was extracted from a conductance map $V_{SD}-V_{g_i}$ where each gate $i$ is scanned over a $\SI{120}{\milli\volt}$ range. The error bars correspond to the uncertainty in determining the gap edge from the conductance maps.}
    \label{fig:all_gates_scan}
\end{figure}

Each data point shown in Figure~4 in the main text was extracted from two distinct Coulomb diamond maps $V_{SD} - V_{g_i}$ $(i = 1, 7)$ at different values of $\Delta V_g$. For each map, one can extract the gap for gate $i$ and then average it over the three gates. The choice of probing only two gates comes from the fact that the gap is a global property of the system that is weakly dependent on the position of the gate.
Supplementary~Figure~\ref{fig:all_gates_scan} shows the gap value for each single gate at an average gate voltage $\braket{V_g} = \SI{-0.2}{\volt}$. All gates exhibit the same global behavior with respect to the dependence of the gap. One can calculate the  standard deviation for each point $\Delta V_g$ which is found to be between $\SI{0.1}{\milli\volt}$ and $\SI{1.2}{\milli\volt}$ with an average standard deviation of $\sim \SI{0.5}{\milli\volt}$. This means that the effect of each gate differs by $15\%$ in average from the mean over all gates.

However, for some values of $\Delta V_g$, the left and right gates ($1-8$ versus $9 - 15$) show some notable differences in their behavior. This is because the disorder configuration in the tube is not necessarily homogeneous in the nanotube. Here, supplementary Figure~\ref{fig:all_gates_scan} shows that the gates $1-8$ behave similarly, whereas the gates $9-15$ sometimes differ from the average, implying that the disorder configuration is probably more sensitive to these gates.

\subsection{Comparison for different gate points}
\begin{figure}[h]
    \centering
    \includegraphics[width=0.72\textwidth]{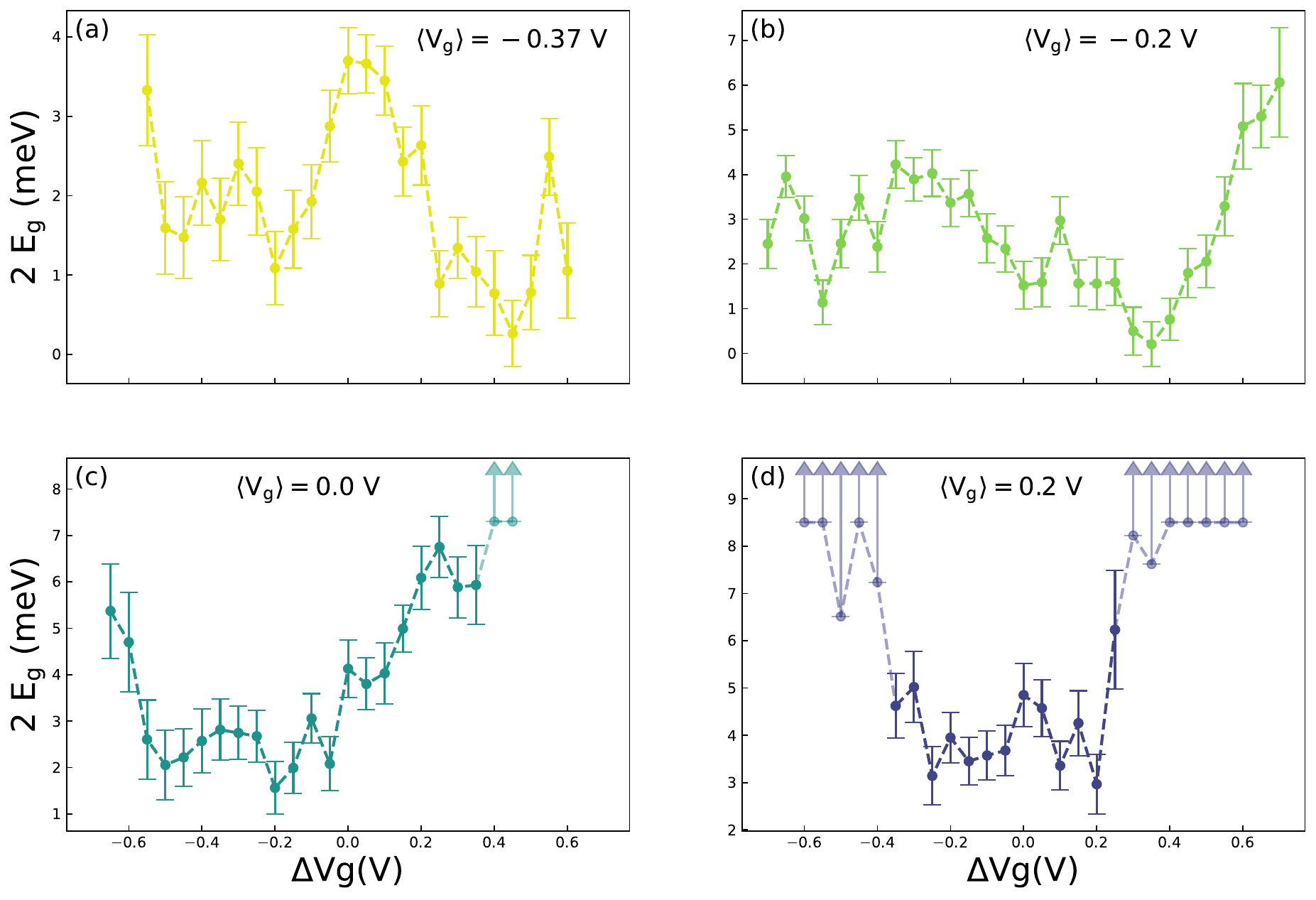}
    \caption{ \textbf{Additional data for the evolution of the gap function for different offset gates.} 
    Gap versus gate modulation $\Delta V_g$ in $V$ from different gate values $\braket{ V_g}$. Each data point was extracted from 3 distinct Coulomb diamond maps $V_{SD} - V_{g_i}$, (gate $i = 1, 7, 14$) at different values of $\Delta V_g$ and averaged over the three gates. Panels (a)-(d) show maximum and minimum values of the gap at different values of $\Delta V_g$ due to the spatial distribution of inhomogeneities being different for each value of $\braket{\Delta V_g}$. The error bars correspond to the uncertainty in determining the gap edge from the conductance maps. }
    \label{fig:compare_Vg_avg}
\end{figure}

The feature shown in the main text in Figure~2 was taken at an average gate point $\braket{V_g} = \SI{0.2}{\volt}$.
For different values of $\braket{V_g}$, the same behavior of a gap closing and opening can be observed as supplementary~Figure~\ref{fig:compare_Vg_avg} shows. All plots show the same qualitative behavior, but with a gap closing at different values of $\Delta V_g$. This shows that disorder may be non-negligible in some gate configurations and can even have a dominant role when compared to the gap induced by the modulation. In the main text, we have chosen to show the dependence of the gap on $\Delta V_g$ for $\braket{V_g} = \SI{0.2}{\volt}$ because the gap was closed for $\Delta V_g \approx \SI{0}{\volt}$. This suggests that for this value $\braket{V_g} = \SI{0.2}{\volt}$, the configuration is less affected by disroder at low $\Delta V_g$ and that the effects of the modulation of $\Delta V_g$ are largely dominant with respect to disorder. 

Some points in Figures~3c. and 3d. are shown with an up arrow signifying that the bias window was not large enough to see the full end of the Coulomb diamonds. For these points, the gap is necessarily larger than the indicated data point.
It is important to note that these data were taken during a previous cool-down of the device compared to the data shown in the main text. This explains the fact that the gap in supplementary Figure~\ref{fig:compare_Vg_avg}.d and in main text (Figures~2 and 4) does not close exactly at the same value of $\Delta V_g$.

Finally, it is important to note that although the curves are different here, they are fully consistent with the larger scale variations shown in Figure 4c of the main text in particular.

\subsection{Chain construction from one side}

Here we present complementary measurements to the chain construction initiated from the center gate, as discussed in the main text. We follow the same experimental protocol, this time starting from one end of the chain, as illustrated in supplementary Figure~\ref{fig:chain_construction_side}. Consistent with previous observations, the modulation-induced gap emerges at finite modulation amplitude $\Delta V_g$ once five or more consecutive gates are modulated. This behavior is further highlighted in supplementary Figure~\ref{fig:chain_construction_gap_vs_sites}, which compares the measured modulation-induced gaps for varying numbers of modulated sites. Panel (a) corresponds to the construction from the center gate, as discussed in the main text, while panel (b) corresponds to the chain construction initiated from the left side. These results demonstrate that a minimum of 5 to 7 modulated sites is necessary to observe the gap-opening behavior, consistently with the Figure~6 of the main text. We can also use this kind of measurement to rule out, like in the main text, an "uncoupled" dot mechanism which would yield a blocking of conductance from a poorly gate coupled spurious quantum dot along the nanotube. 

\begin{figure}[!h]
\centering
\includegraphics[width=1\textwidth]{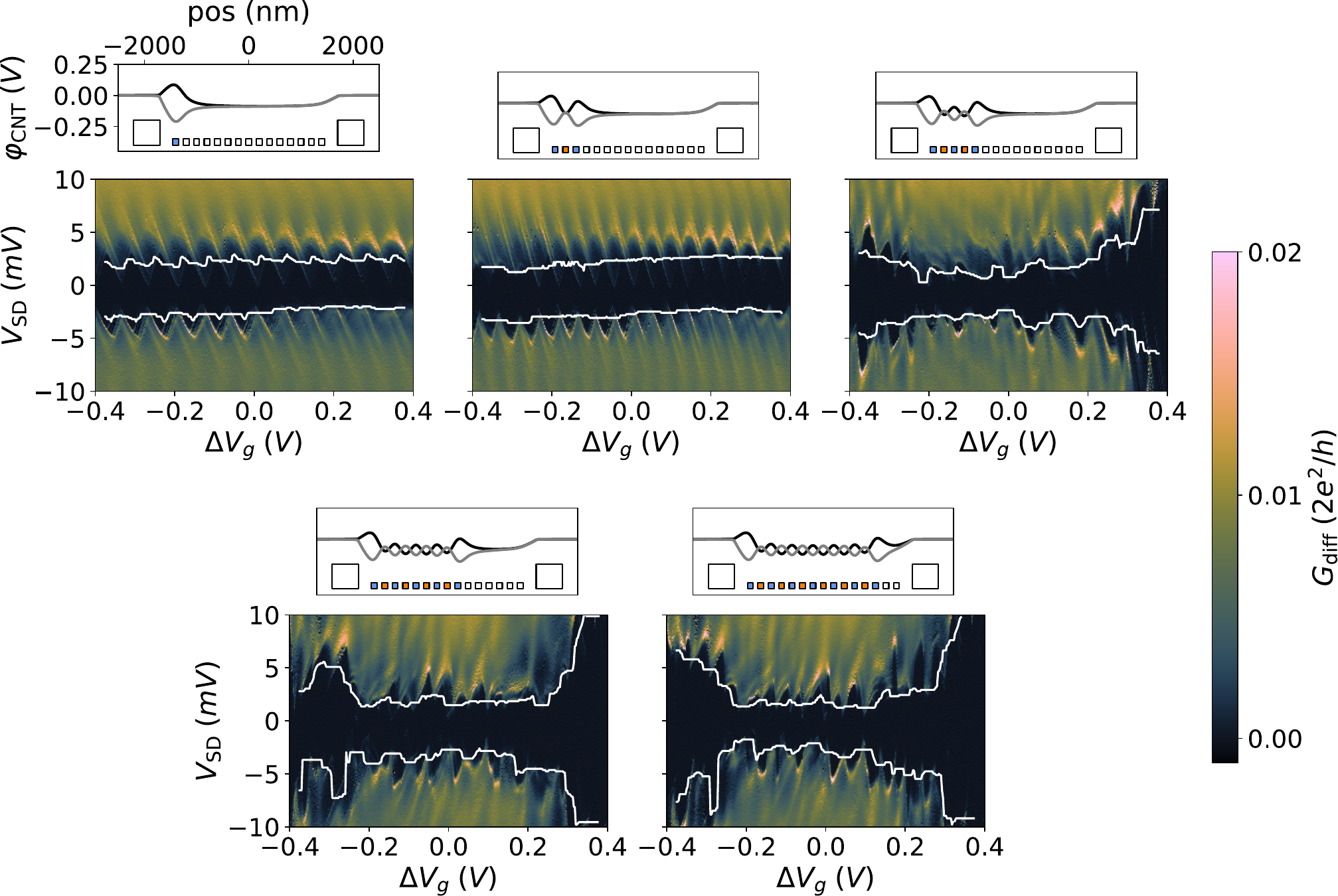}
\caption{\textbf{Chain construction initiated from the left side.} 
Each panel displays the differential conductance $G_{diff}$ as a function of gate modulation amplitude $\Delta V_g$ and bias voltage $V_{SD}$. The white lines indicate the gap threshold. Above each conductance plot, a schematic illustrates modulated gates in blue and orange (alternating gates), while white squares indicate non-modulated gates. Large squares at the edges represent source and drain contacts. Grey and black curves show the calculated electrostatic potential profile along the carbon nanotube for the two extreme modulation amplitudes.}
\label{fig:chain_construction_side}
\end{figure}

\begin{figure}[!h]
\centering
\includegraphics[width=0.8\textwidth]{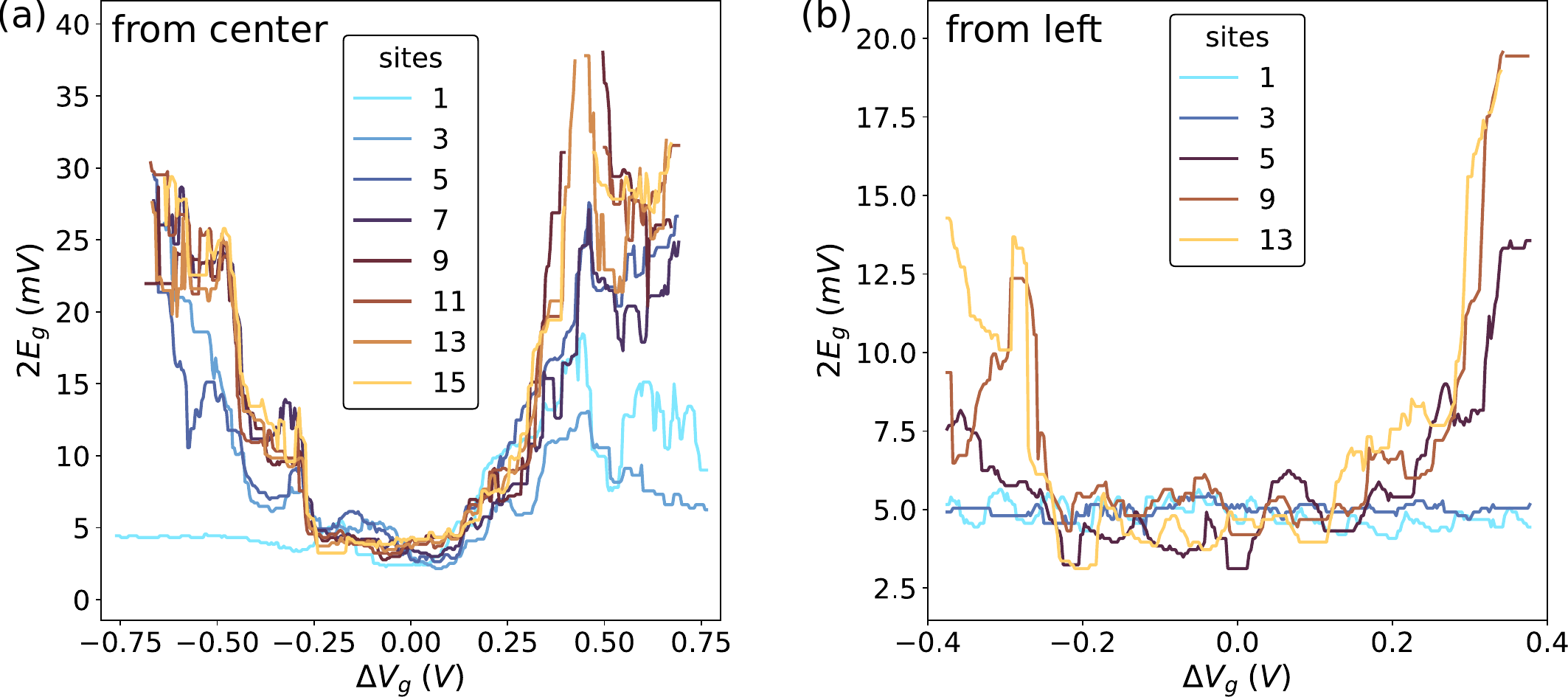}
\caption{\textbf{Modulation-induced gap as a function of gate modulation amplitude $\Delta V_g$ for varying numbers of modulated sites.} 
(a) Data for chain construction initiated from the center gate, as presented in the main text. (b) Data for chain construction initiated from the left side, corresponding to supplementary Figure~\ref{fig:chain_construction_side}.}
\label{fig:chain_construction_gap_vs_sites}
\end{figure}

\section{Gate dependence of current at room temperature and at low temperatures}

\begin{figure}[!h]
\centering
\includegraphics[width=0.8\textwidth]{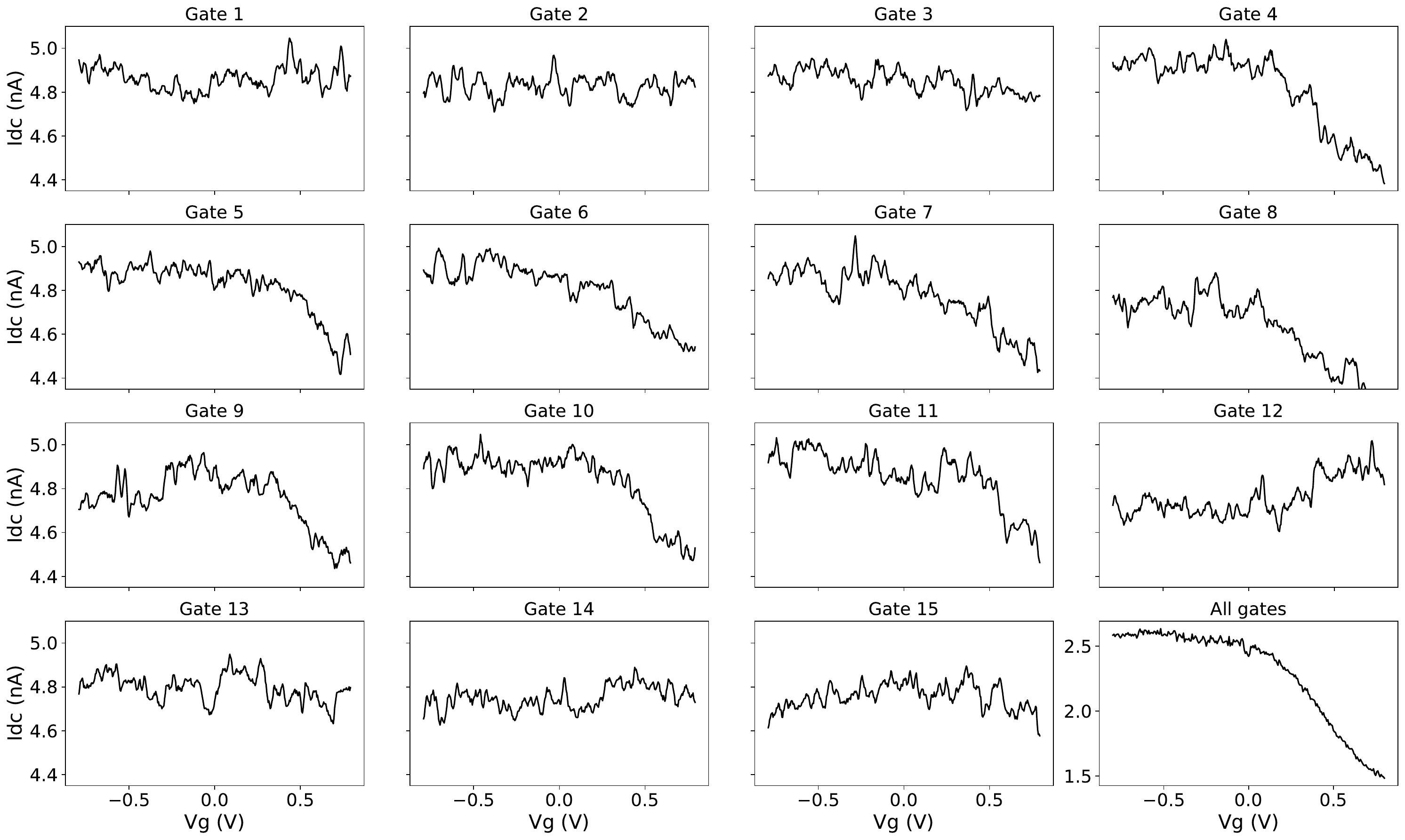}
\caption{\textbf{Room temperature characterization.} The 15 first panels correspond to each Gate $i\in \{1-15\}$ response for each gate at room temperature. The last panel is the 'all gates' response.}
\label{fig:IvgRT}
\end{figure}

We present here the gate dependence at room and low temperature for the second cooldown. In Figure~\ref{fig:IvgRT}, we present the gate response at room temperature of the device for the second cool down. We see only a weak dependence of the current at $V_{SD}=$ \SI{2}{\milli\electronvolt} for each gate. For gates 1,2,3 and 13,14,15, there is essentially no gate dependence whereas for the gates in the middle, there is a $10\%$ variation. This shows a rather spatially homogeneous response of the nanotube to the local gates. The weak dependence is attributed to the screening of the gates with each other. When all gates are swept, the current displays a down turn around \SI{0.5}{\volt} signaling semiconducting behavior 
with a gap above \SI{0.5}{\volt}. We have not measured way above this value to prevent the nanotube to collapse on the gate keyboard. A low temperatures (\SI{30}{\milli\kelvin}), the current is not measurable above about \SI{0.1}{\volt}, indicating a semiconducting gap. Note that our measurements are performed away from the bottom of the band, at the scale of the Coulomb diamonds.

\begin{figure}[!h]
\centering
\includegraphics[width=0.8\textwidth]{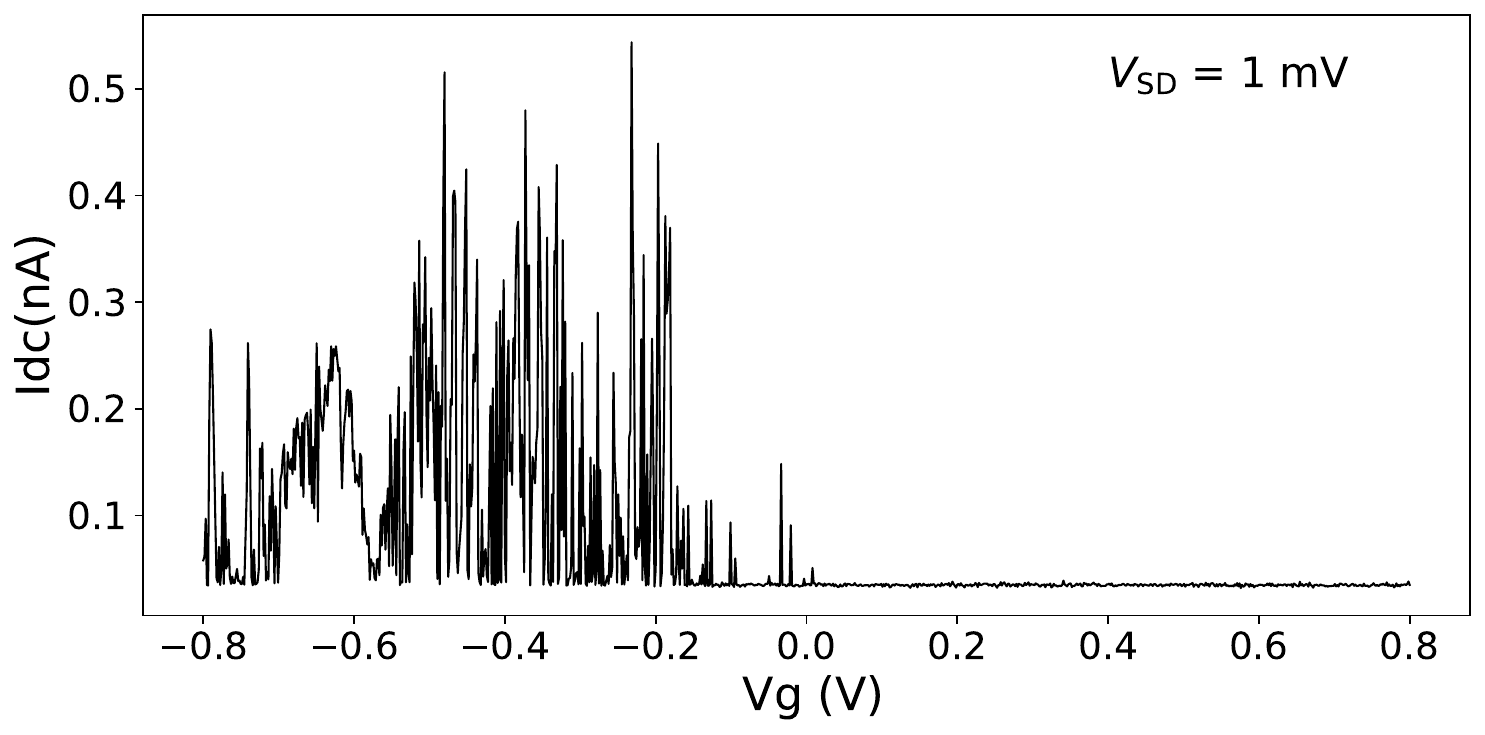}
\caption{\textbf{Gate response at low temperature. } 
Conductance of the device as a function of the gate voltage at $20mK$}
\label{fig:IvgCryoT}
\end{figure}
\clearpage

\section{Fabrication methods}
The device shown in the main text, in Figure~1, is a superconducting circuit fabricated from an undoped $\SI{500}{\micro\metre}$ silicon wafer covered with $\SI{500}{\nano\metre}$ of $\mathrm{SiO_2}$. To make the ground plane, the gates, DC lines, and the contact electrodes, different metals are deposited by electron beam evaporation. We use for the layout standard e-beam lithography techniques. At the end of these processes, a carbon nanotube is deposited on top of the contact electrodes ensuring that it is not polluted by any nanofabrication residue.

\subsection{Circuit fabrication}
The circuit consists of the following elements:
\begin{itemize}
    \item 15 DC gates (12 DC only, 2 RF + DC and 1 cavity gate)
    \item 1 microwave CPW cavity that also gates the tube
    \item 2 electrodes (source and drain) + 2 electrodes used to cut the nanotube
    \item 2 trenches to enable the CNT deposition
\end{itemize}

The DC lines, RF lines, ground plane and cavity are etched by reactive ion etching (RIE) after a full chip metalization of \SI{90}{\nano\metre} of niobium. The cavity is a centimetric CPW resonator $L = \lambda/2$. The impedance of all the DC and RF lines is $\SI{50}{\ohm}$ to avoid reflections with external RF and DC sources.
The 15 DC gates are \SI{100}{\nano\metre} wide and spaced by \SI{100}{\nano\metre} from each other and are made of a stack of \SI{30}{\nano\metre} Ti / \SI{25}{\nano\metre} Pd. The electrodes are made of a stack of \SI{200}{\nano\metre} Ti / \SI{25}{\nano\metre} Pd. The gate-CNT distance of \SI{170}{\nano\metre} is chosen so that the measurements are not sensitive to the spatial displacement of the CNT \cite{waissman_realization_2013}. In addition, if the gates were too close, it would lead to a screening of the electron-electron interactions. In return, a large distance induces a non-local gating effect on the electrons in the CNT. Finally, since the tube extends over a long distance ($\sim \SI{4}{\micro\metre}$), the electrodes must be high enough so that the tube does not collapse onto the gates.

To host the cantilevers, necessary for the CNT deposition, two trenches are etched from silicon on both sides of the electrodes. These trenches are $\sim \SI{15}{\micro\metre}$ deep and are etched by reactive ion etching while the rest of the chip is protected by a thick $\sim \SI{3}{\micro\metre}$ resist.

\subsection{Carbon nanotube deposition}
The CNTs are grown on silicon cantilevers by $\mathrm{CH_4}$ chemical vapor deposition. The nanotubes can also be electrically characterized and imaged using a scanning electron microscope.
Once the circuit is fabricated, a nanotube is deposited on top of the contact electrodes under vacuum through a nanoscale transfer method \cite{viennot_stamping_2014, gramich_fork_2015, blien_quartz_2018}. This process allows us to select a clean CNT that never undergoes any clean-room process.

A CNT is approached close to the chip surface while a \SI{2}{\volt} DC bias is applied between the contact electrodes under a high $10^{-7}\SI{}{\milli\bar}$ vacuum. When the nanotube reaches the surface of the electrode, a current is measured, allowing us to detect it electrically. Before removing the comb, the tube is cut on both sides by applying a high current between the lead and the cut electrode (typically $5-15\SI{}{\micro\ampere}$). This last step also acts as an annealing procedure. Finally, one can measure the source-drain resistance that typically reaches \SI{500}{\kilo\ohm}.

\section{Mathieu equation details}\label{section:mathieu}

To better understand the appearance of a gap when electrons are subject to a dimerized potential $V_g = \braket{V_g}\pm\Delta V_g$, one would need to solve the "Poisson-Schr\"odinger" problem with such a gate configuration. A much easier problem to solve is to consider, first, the electrical potential induced by the gates in the absence of other charges, and then, to solve the Schr\"odinger equation in the nanotube subject to such a potential.

\begin{figure}[h]
    \centering
    \includegraphics[width=0.6\textwidth]{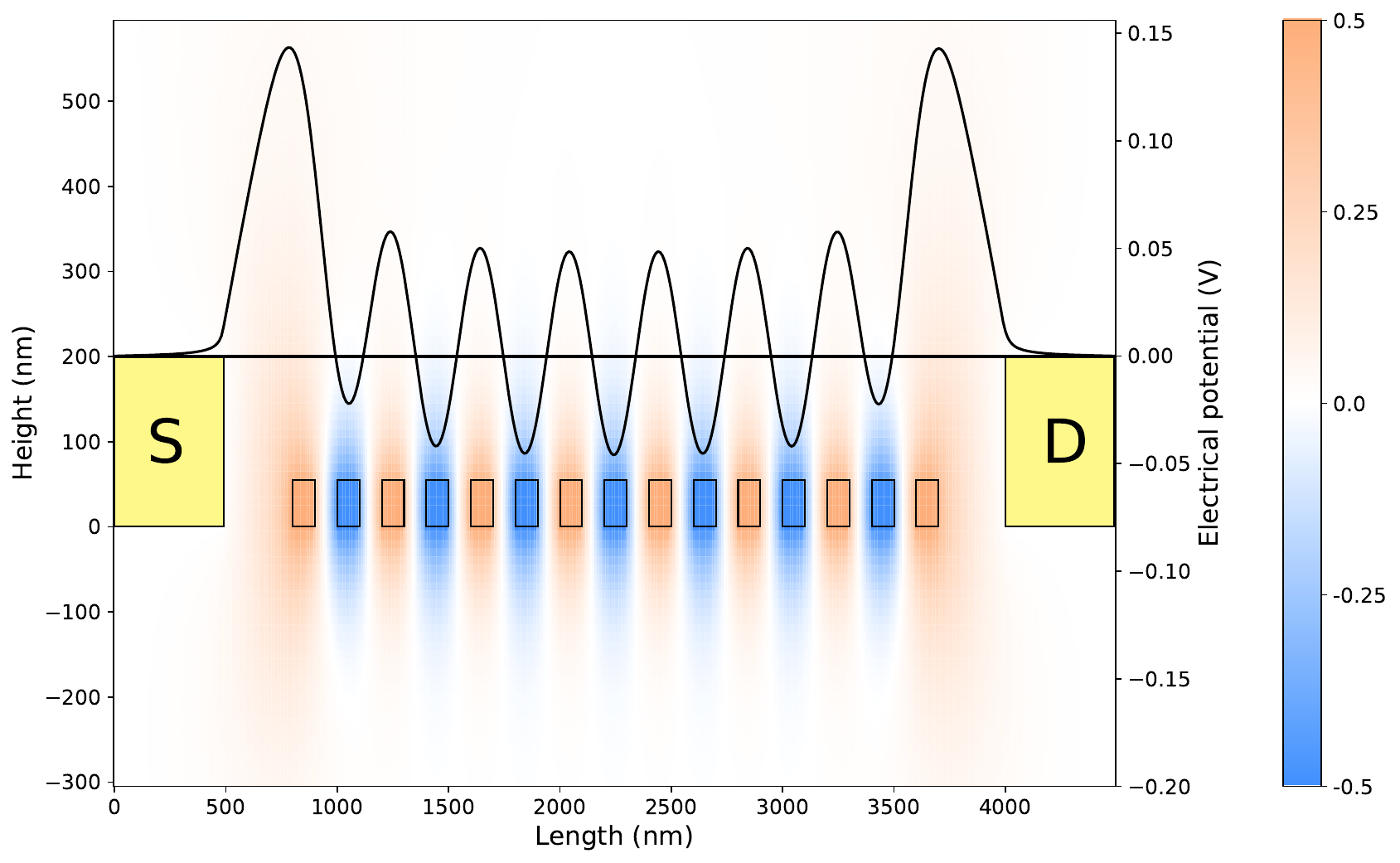}
    \caption{\textbf{Potential landscape.} 
    Calculated electrostatic potential using the Poisson equation. The curve shows the potential at the CNT location.}
    \label{fig:poisson}
\end{figure}

The potential induced by 15 gates in free space with gate values $V_g = \braket{V_g}\pm\Delta V_g$ is expected to be sinusoidal. Figure 4 shows the numerical solution of the electrical potential induced by 15 gates with potential $\pm \SI{0.5}{\volt}$. It corresponds to a case where $\braket{V_g} = 0$ and $\Delta V_g = \SI{0.5}{\volt}$. In the inner part of the system, we recover a sinusoidal function as expected. For the central gates, the screening is rather homogeneous, whereas the edge electrodes have a different electrostatic environment. An important point to note is that in the center of the tube, the amplitude of the potential is $\approx \SI{50}{\milli\volt} $, which is in good agreement with the lever arm measured of $\alpha \approx 0.12$.

\subsection{Analytical approach}

Considering a perfect cosine function for the potential, one can write the 1D, time-independent Schrödinger equation as :
\begin{equation} \label{eq:schro_cos}
     -\frac{\hbar^2}{2m_{eff}}\frac{\partial^2\Psi}{\partial x^2} + V_0\cos\left(\frac{2\pi x}{\lambda_{mod}}\right)\Psi = E\Psi
\end{equation}

Where $m_{eff}$ is the effective mass of electrons in the nanotube, $V_0$ is the amplitude of the modulation and $\lambda_{mod}$ is the wave length of the modulation.
This equation can be rewritten as the Mathieu's equation:

\begin{equation} \label{eq:mathieu}
    \frac{\partial^2\Psi}{\partial z^2} + \left(a - 2\eta \right)\cos\left(2z\right)\Psi=0
\end{equation}

with $z = \frac{\pi x}{\lambda_{mod}}$, $\eta = \frac{V_0}{2\delta}$, $\delta = \frac{\hbar^2\pi^2}{2m_{eff}\lambda^2_{mod}}$ and $a = \frac{E}{\delta}$.

This equation can be solved asymptotically, and thus the band gap can be computed numerically. At small $\eta$, the first band gap is found to be $E_g/\delta = 2\eta$ (up to the second order in $\eta$). And for large $\eta$, it is equal to $E_g/\delta \approx 4\sqrt{\eta} $.

Since the gates are $100 nm$ wide and spaced by $100 nm$, the modulation wavelength is found to be $\lambda_{mod} = 400 nm$. Taking into account a semiconducting gap in the nanotube of $\Delta_g \simeq 80 meV$ and a Fermi velocity $v_f \simeq 8.10^5 m.s^{-1}$, the effective mass of the electrons can be calculated as $m_{eff} \sim \frac{\Delta_g}{v_f^2} \simeq 2.10^{-32} kg$. $\delta \simeq 100 \mu eV$. The amplitude of the gate modulation is typically a few hundred $mV$ up to a lever arm. The slopes of the Coulomb blockade peaks allow us to extract a lever arm $\alpha = 0.12$. And then, we can compute that $V_0 \gg 1 meV$.
In this case, $V_0 \gg \delta$ and thus the modulation gap is
\begin{equation} \label{eq:gap_mathieu}
E_g \approx 2\sqrt{2\delta \alpha \lvert \Delta V_g \rvert}
\end{equation}

The square root dispersion of the modulation gap is still valid for higher index bands but there is a threshold increasing with the band index for the onset of the modulations gap, as shown in Figure 3 of the main text. The effect of disorder which is the main competing mechanism is discussed in the next section.

\subsection{Numerical approach}\label{ssec:numerical_mathieu}

We numerically solve the equivalent Mathieu problem described by eq.~\eqref{eq:schro_cos}. We consider a one-dimensional spatial region with length equal to the period of the cosine electrostatic potential, $\lambda_{mod}$. The region is discretized into $N = \lambda_{mod} / \mathrm{dx}$ equally spaced points, where $\mathrm{dx}$ is the spatial step size. The Hamiltonian matrix is given by:

\begin{equation}\label{eq:Schrodinger_periodic_H}
H(k) \;=\;
\begin{bmatrix}
2t+V_{1} & -t & 0 & \cdots & -t\,e^{-ika} \\[4pt]
-t & 2t+V_{2} & -t & \ddots & 0 \\[4pt]
0 & -t & 2t+V_{3} & \ddots & 0 \\[4pt]
\vdots & \ddots & \ddots & \ddots & -t \\[4pt]
-t\,e^{+ika} & 0 & 0 & -t & 2t+V_{N}
\end{bmatrix},
\end{equation}

with $V_i = V_0 \cos \left[ \frac{2 \pi (i - 1)}{N} \right]$, $i \in [1, N]$, hopping parameter $t = \frac{\hbar^{2}}{2m_{eff}(\mathrm{dx})^{2}}$, and wave  vector $k \in \bigl[-\frac{\pi}{\lambda_{mod}},\,\frac{\pi}{\lambda_{mod}}\bigr]$. The phase factors $e^{\pm ika}$ coupling the first and last sites arise from Bloch's theorem, which implies periodic boundary conditions: $\Psi(x) = e^{-ika}\Psi(x+a)$ for plane-wave solutions $\Psi(x)$.

We diagonalize this Hamiltonian for various values of $V_0$ and $k$ to obtain the energy bands and the associated gaps in the Mathieu problem, as illustrated in the main text. In practice, the Hamiltonian is solved explicitly for two critical points $k=0$ and $k=\pi$. The gap between successive bands $n$ and $n+1$ is determined as $2 E_g(n) = 2\left(E_{n+1}(k_n) - E_{n}(k_n)\right)$, where $k_n$ is either $0$ or $\pi$, depending on the band pair under consideration.

Additionally, we incorporate disorder into the system by assigning to each lattice site a randomly distributed potential. This potential is drawn from a normal distribution centered at zero, with standard deviation $\sigma_{dis}$.

For numerical simulations, we use parameters: $N = 400$, $\mathrm{dx} = \SI{1}{\nano\metre}$ (giving $\lambda_{mod} = \SI{400}{\nano\metre}$), semiconducting gap $E_0 = \SI{40}{\milli\electronvolt}$, and Fermi velocity $v_F = 8 \times 10^5\,\SI{}{\metre.\second^{-1}}$, which yields an effective mass $m_{eff} = 0.011 m_e$, where $m_e = 9.1 \times 10^{-31}\,\SI{}{\kilo\gram}$ is the electron mass. We take an electrostatic lever arm of $\alpha=0.25$ to convert the gate modulation voltage to energy, $V_0 = \alpha \Delta V_g$.

We found that introducing disorder with $\sigma_{dis} = \SI{0.015}{\electronvolt}$ reproduces the experimentally observed gap value of approximately $\SI{2}{\milli\electronvolt}$ at low modulation amplitudes $\Delta V_g$. We perform simulations over 400 disorder realizations, computing both the mean and standard deviation of $2 E_g$ at each modulation amplitude. Mean values and standard deviations are depicted as dashed lines and shaded areas, respectively. Individual disorder realizations, which illustrate substantial variation in gap magnitude at small modulation amplitudes, are shown as solid lines.

A specific dispersion of the energy bands versus modulation amplitude $\Delta V_g$ is presented in supplementary Figure~\ref{fig:bands_with_disorder}. This Figure~highlights deviations from the ideal cosine potential and shows the extraction of modulation-induced band gaps.

\begin{figure}[h]
\centering
\includegraphics[width=0.5\textwidth]{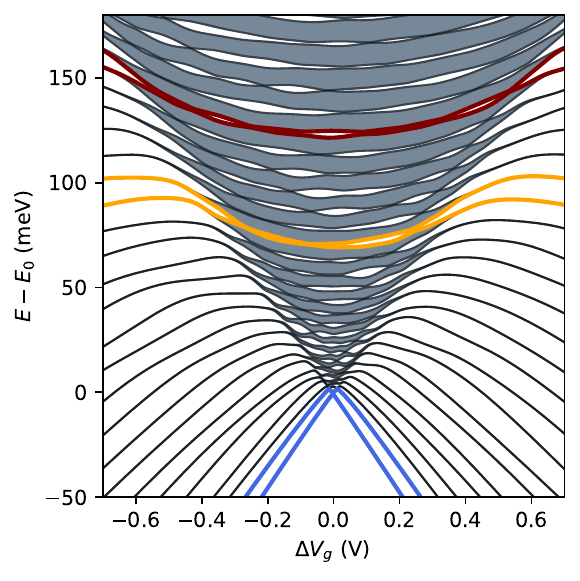}
\caption{\textbf{Effect of disorder.} 
Energy bands in the presence of an infinite cosine electrostatic potential with disorder of amplitude $\sigma_{dis} = \SI{0.015}{\electronvolt}$, plotted against modulation amplitude $\Delta V_g$. The blue, orange, and red curves emphasize the band gap openings at energy or chemical potentials measured from the band bottom.}
\label{fig:bands_with_disorder}
\end{figure}